\documentclass[paper]{JHEP3}
\usepackage{times}
\usepackage{amsmath, amsthm, amssymb}
\usepackage{mathrsfs}
\usepackage[dvips]{graphicx}
\usepackage{picinpar}
\usepackage{slashed}
\usepackage{booktabs}
\usepackage{cite}
\newcommand{\eV}{\, \textrm{eV}}

\newcommand{\GeV}{\, \textrm{GeV}}

\newcommand{\myE}{\mathcal{E}}
 
\title{Wash-Out in \boldmath{$N_2$}-dominated leptogenesis}

\author{{\slshape F.~Hahn-Woernle$\,^1$}\\
  $^1$Max-Planck-Institut f\"ur Physik, F\"ohringer Ring 6,
  D-80805 M\"unchen, Germany\\ \texttt{fhahnwo@mppmu.mpg.de}}

\abstract{We study the wash-out of a cosmological baryon asymmetry
  produced via leptogenesis by subsequent interactions.  Therefore we
  focus on a scenario in which a lepton asymmetry is established in
  the out-of-equilibrium decays of the next-to-lightest right-handed
  neutrino. We apply the full classical Boltzmann equations without
  the assumption of kinetic equilibrium and including all quantum
  statistical factors to calculate the wash-out of the lepton
  asymmetry by interactions of the lightest right-handed state. We
  include scattering processes with top quarks in our analysis. This
  is of particular interest since the wash-out is enhanced by
  scatterings and the use of mode equations with quantum statistical
  distribution functions. In this way we provide a restriction on the
  parameter space for this scenario.}

\preprint{MPP-2009-209\\
 }

\begin{document}

\section{Introduction and conventions}
\label{sec:introduction}
Relying on the see-saw
mechanism~\cite{Minkowski:1977sc,Yanagida:seesaw,Gell-Man:sugra,Barbieri:1979ag,Mohapatra:1980yp}
to create small masses for the Standard Model (SM) neutrinos,
leptogenesis~\cite{Fukugita:1986hr} provides an attractive explanation
of the baryon asymmetry of the universe. The main ingredients of
leptogenesis are the following: the out-of-equilibrium decays of the
heavy right-handed neutrinos, that were added in the see-saw
mechanism, into leptons and Higgs particles violate $CP$, from which a
lepton asymmetry can be generated.  This lepton asymmetry is then
partially transformed into a baryon asymmetry by anomalous processes
of the SM called sphalerons~\cite{Klinkhamer:1984di}. In this way the
three Sakharov conditions are fulfilled and leptogenesis turns out as
a consequence of the see-saw mechanism.

In the minimal (type-I) see-saw model, gauge singlet fermions with
Majorana masses $M_i$, eigenvalues of the complex symmetric $3 \times
3$ Majorana mass matrix $M_{M}$, couple to the massless lepton doublet
and to the Higgs doublet of the SM through Yukawa couplings given by
the $3 \times 3$ matrix $\lambda_{\nu}$. These Yukawa couplings then
generate a Dirac mass term $m_{D}=\lambda \,v$, where $v=174\GeV$ is
the vacuum expectation value of the Higgs field, relating the heavy
singlets to SM neutrinos upon spontaneous symmetry breaking of the
electroweak gauge symmetry. In the see-saw mechanism the
eigenvalues of the Majorana mass matrix are much larger than the Dirac
mass matrix entries, $M_{M_{i}} \gg m_{D_{i}}$, and the mass matrix of
the left-handed states is then suppressed by the high-energy scale
\textit{M},
\begin{align}
\label{eq:m-sesa}
m_{\nu_{ij}}=-v^2\lambda_{ik}M_{M_k}^{-1}\lambda_{kj}.
\end{align}
Participating in Yukawa couplings, the heavy neutrinos are unstable
and the total decay width of a right-handed neutrino generation $N_i$
at tree-level is given by
\begin{align}
  \label{eq:decay-family}
  \Gamma_{D_{i}} =\sum\limits_{\alpha} \left( \Gamma_{i\alpha} (N_{i}
    \rightarrow \Phi+l_{\alpha}) + \overline{\Gamma}_{i\alpha} (N_{i}
    \rightarrow \overline{\Phi} + \overline{l}_{\alpha})\right) =
  \frac{1}{8 \pi} \left( \lambda^{\dagger} \lambda \right)_{ii} M_{i},
\end{align}
where the sum is taken over the single decay rates into lepton flavor
$\alpha$. The CP asymmetry in lepton flavor $\alpha$ is then defined
as
\begin{align}
  \label{eq:asy-general}
  \varepsilon_{i \alpha} \equiv \frac{\Gamma_{i\alpha} (N_{i}
    \rightarrow \Phi+l_{\alpha}) - \overline{\Gamma}_{i\alpha} (N_{i}
    \rightarrow \overline{\Phi}+\overline{l}_{\alpha})}
  {\sum_{\alpha} \left(\Gamma_{i\alpha} (N_{i} \rightarrow
      \Phi+l_{\alpha}) + \overline{\Gamma}_{i\alpha} ( N_{i} \rightarrow
      \overline{\Phi} + \overline{l}_{\alpha})\right)},
\end{align}
This asymmetry is due to the interference of the tree-level amplitude
and the one-loop vertex and self-energy contributions. The indices $i$
and $\alpha$ denote the generation of the decaying right-handed
neutrino and the flavor of the produced lepton,
respectively. Accounting for different lepton generations, the $CP$
asymmetry is a diagonal matrix in flavor space.  It has been shown
in~\cite{Abada:2006fw,Abada:2006ea,Nardi:2006fx,Blanchet:2006be,Blanchet:2006ch}
that the flavor structure of the final state leptons has sizable
implications on leptogenesis predictions. To account for the lepton
flavor structure one introduces the projectors
\begin{align}
\label{eq:fl-proj-1}
P_{i\alpha}&\equiv \vert \langle l_i\vert l_{\alpha} \rangle \vert^2 =
P^0_{i\alpha} + \frac{\Delta P^0_{i\alpha}}{2}, \\ \label{eq:fl-proj-2}
\overline{P}_{i\alpha}&\equiv \vert \langle
\overline{l}^{\prime}_i\vert \overline{l}_{\alpha} \rangle \vert^2 =
P^0_{i\alpha} - \frac{\Delta P^0_{i\alpha}}{2}.
\end{align}
The wash-out rate in inverse decays is then reduced by the projector
$P^0_{i\alpha}=\left( P_{i\alpha}+\overline{P}_{i\alpha} \right)/2$,
since the Higgs will make inverse decays on flavor eigenstates $\vert
l_{\alpha}\rangle$ instead of the linear superposition $\vert
l_i\rangle$. The second effect, an additional $CP$-violating
contribution, stemming from a different flavor composition between
$\vert l_i\rangle$ and $CP\,\vert \overline{l}^{\prime}_i\rangle$, can
be expressed by the projector difference $\Delta P_{i\alpha}=
P_{i\alpha}-\overline{P}_{i\alpha}$. The sum over flavor of this
difference results in $\sum_{\alpha}\,\Delta P^0_{i\alpha}=0$. The
decay rates into single lepton flavor $\alpha$ are now given by
$\Gamma_{i\alpha}\equiv P_{i\alpha}\Gamma_i$ and
$\overline{\Gamma}_{i\alpha}\equiv
\overline{P}_{i\alpha}\overline{\Gamma}_i$, and the flavored $CP$
asymmetries can be written in terms of the unflavored asymmetries
\begin{align}
\label{eq:asy-general-1}
\varepsilon_{i\alpha}= P^{0}_{i\alpha}\,\varepsilon_i+\frac{\Delta
  P_{i\alpha}}{2}.
\end{align} 
The second term gives the contribution due to a potentially different
flavor composition between $\vert l_i\rangle$ and
$\vert\overline{l}^{\prime}_i\rangle$. Summing these $CP$ asymmetries
over the flavor yields the total $CP$ asymmetry used in the unflavored
regime $\varepsilon_i=\sum_{\alpha} \varepsilon_{i\alpha}$. In this
paper we will mainly be interested in the effects of a full spectral
distribution of right-handed neutrinos on leptogenesis and not in the
effects of the lepton flavor structure. Therefore we will perform all
calculations in the so-called alignment case, that is realized when
the $N_i$-decays are just into one flavor $\alpha$, resulting in
$P_{i\alpha}=\overline{P}_{i\alpha}=1$,
$P_{i\beta\neq\alpha}=\overline{P}_{i\beta\neq\alpha}=0$ and
$\varepsilon_{i\alpha}=\varepsilon_{i}$. In this case the kinetic
equations, c.f. Section~\ref{subsec:FullBE}, are identical to the
kinetic equations used in the unflavored regime that is realized at
$T\gtrsim 10^{12}\GeV$. By restricting to the alignment case we can
compare our results to the ones obtained in previous calculations
using unflavored Boltzmann
equations~\cite{DiBari:2005st,Garayoa:2009my,HahnWoernle:2009qn}. However,
we will keep all flavor indices in our notation, since all results are
valid when considering independently the single flavors $\alpha$.  We
will comment on the effects of the lepton flavor structure in more
detail later in Section~\ref{sec:NoteFlavor}.

Summing over all lepton flavors, $\varepsilon_i=\sum\limits_{\alpha}
\varepsilon_{i\alpha}$, the CP asymmetry can be written
as~\cite{Flanz:1994yx,Covi:1996wh,Buchmuller:2000as}
\begin{align}
\label{eq:CP-asy-n2}
\varepsilon_i \approx - \frac{1}{8 \pi} \sum\limits_{\substack{j=1,2,3
    \\ j\neq i}} \frac{\textrm{Im} \left[ \left( \lambda^{\dagger}
      \lambda \right)^2_{ij} \right]}{\left( \lambda^{\dagger}\lambda
  \right)_{ii}} \times \left[ f_V \left( \frac{M^2_j}{M^2_i} \right) +
  f_S \left( \frac{M^2_j}{M^2_i} \right) \right],
\end{align} 
where the vertex ($f_V)$ and the self-energy ($f_S$) contributions are
\begin{align}
  \label{eq:cp-contributions}
  f_V(x)= \sqrt{x}\left[ 1-\left( 1+x \right)\,\log \left(
      \frac{1+x}{x} \right) \right], \quad\textrm{and}\quad
  f_S(x)=\frac{\sqrt{x}}{1-x}.
\end{align}
The out-of-equilibrium dynamic that is necessary for successful
leptogenesis is provided by the expansion of the universe. One usually
compares the total decay rate of the right-handed neutrino state given
in Eq.~(\ref{eq:decay-family}) to the expansion rate at temperatures
$T\sim M_i$,
\begin{align}
\label{eq:exp-rate}
H(T=M_i)=\sqrt{4\pi^{3}g^{\ast}/45}\left(M_i/M_{\rm Pl}\right),
\end{align}
introducing the decay parameters
\begin{align}
  \label{eq:K-1}
  K_{i}\equiv\frac{\Gamma_{D_{i}}}{H(M_{i})}=\frac{\tilde{m}_{i}}{m_{\ast}},
\end{align}
where $M_{\rm Pl}=1.221 \times 10^{19} \ {\rm GeV}$ is the Planck
mass, and $g^{\ast}=106.75$ corresponds to the number of relativistic
degrees of freedom in the SM at temperatures higher than the
electroweak scale.  One introduces at this point two dimensionful
variables to connect the decay parameters to the neutrino mass scale
via the {\it effective neutrino mass}~\cite{Plumacher:1996kc}
$\tilde{m}_i$ and the {\it equilibrium neutrino mass} $m_{\ast}$:
\begin{align}
  \label{eq:mu-nu-eff-eq}
  \tilde{m}_{i}=\frac{v^2(\lambda^{\dagger}\lambda)_{ii}}{M_i},
  \quad\textrm{and}\quad   m_{\ast}=\frac{16\pi^{\frac{5}{2}}\sqrt{g^{\ast}}}
  {3\sqrt{5}}\frac{v^{2}}{M_{\rm{Pl}}}\approx 1.08\times 10^{-3} \eV.
\end{align}
The decay parameters for each flavor $\alpha$ are again given via the
projectors 
\begin{align}
\label{eq:K-alpha}
K_{i\alpha}\equiv P^0_{i\alpha}\,K_i.
\end{align}
The decay parameters control whether the right-handed neutrino decays
in equilibrium \mbox{($K_{i\alpha}>1$)} or out of equilibrium
($K_{i\alpha}<1$) and are key quantities for the dynamics of
leptogenesis.

In general, the baryon asymmetry produced by leptogenesis can be
written as~\cite{Buchmuller:2004nz}
\begin{align}
  \label{eq:ba-asy}
  \eta_B=\frac{3}{4}\frac{\alpha_{\rm{sph}}}{f}\,\sum\limits_{i,
    \alpha} \varepsilon_{i\alpha} \,\kappa^{f}_{i\alpha}\equiv
  d\,\sum\limits_{i,\alpha}\varepsilon_{i\alpha}\,\kappa^{f}_{i\alpha}
  \simeq 0.96\times
  10^{-2}\,\sum\limits_i\varepsilon_i\,\kappa^{f}_{i},
\end{align}
where in the last step the sum over the flavor $\alpha$ has been
performed as it is possible in the alignment case or in the unflavored
regime at $T\gtrsim 10^{12}\GeV$. The final efficiency factors
$\kappa^{f}_{i}$, that parametrize the amount of asymmetry that survives
the competing production and wash-out processes are direct results of
solving the relevant Boltzmann equations for
leptogenesis~\cite{Plumacher:1997ru,Buchmuller:1996pa}.  In the limit
of vanishing wash-out and a thermal initial abundance for the
right-handed neutrinos, the efficiency factors have a final value
$\kappa^{f}_{i}=1$. The factor $f=2387/86$ accounts for the dilution of
the baryon asymmetry due to photon production from the onset of
leptogenesis till recombination and the quantity
$\alpha_{\textrm{sph}} =28/79$ is the conversion factor of the $B-L$
asymmetry into a baryon asymmetry by the sphaleron processes.

The most simple realization of thermal leptogenesis is the
$N_1$-dominated scenario, where a lepton asymmetry is exclusively
generated in the decays of the lightest right-handed state
$N_{1}$. Qualitatively, this scenario can be realized assuming $M_1
\ll M_2 \ll 10^{14}\GeV \ll M_3$. Then the heaviest right-handed state
$N_3$ decouples~\cite{Chankowski:2003rr} and this typically implies a
hierarchy in the $CP$ asymmetries, i.e., $\vert \varepsilon_{3} \vert
\ll \vert \varepsilon_{2} \vert \ll \vert \varepsilon_1 \vert$, as a
consequence of light particles in the internal lines of the self-energy
and vertex corrections to the tree-level decay diagram. Indeed, in the
limit of massless particles in the internal lines, the corresponding
$CP$ asymmetry vanishes, cf. Eq.~(\ref{eq:CP-asy-n2}) . Furthermore,
due to $N_1$ interactions following the decays of the heavier
right-handed states $N_{2,3}$, a substantial part of the produced
lepton asymmetry will presumably be washed-out, especially in the
regime of strong wash-out.

In this paper we will go beyond the simplest scenarios and consider
leptogenesis in a set-up where the asymmetry is generated in the
decays of the next-to-lightest state $N_2$, that is called the
$N_2$-dominated scenario~\cite{DiBari:2005st}.  The viability of
$N_2$-dominated leptogenesis is restricted by the wash-out of produced
asymmetry induced by interactions of the lightest state $N_1$ that
tend to destroy any previously created lepton asymmetry. We will address
this issue of wash-out in $N_2$-dominated leptogenesis within the
simple alignment case, that is equivalent to the unflavored regime, by
means of complete Boltzmann equations at the mode level including
scattering with top quarks\cite{HahnWoernle:2009qn} and extend on the
previous studies of\cite{Garayoa:2009my} where only decays and inverse
decays were considered. We show how the additional wash-out due to
quantum statistical distribution functions and scattering processes
restricts the valid parameter space of $N_2$-dominated leptogenesis.

In the next section we show how specific scenarios of leptogenesis can
be realized by specifying the relevant see-saw parameters and
discuss qualitatively the $N_1$- and $N_2$-dominated scenarios. In
Section~\ref{subsec:FullBE} we present Boltzmann equations for
leptogenesis including scattering at the mode level and apply these
equations to study $N_2$-dominated leptogenesis quantitatively in
Section~\ref{subsec:ful-n2domination}. In the
Appendices~\ref{subsec:full-s-channel},~\ref{subsec:full-t-channel},
and~\ref{subsec:reacrates} we list explicitly the integral
expressions used to solve the Boltzmann equations.

\section{The orthogonal matrix and different scenarios
  of leptogenesis}
\label{subsec:n2-realization}
To understand how specific scenarios of leptogenesis can be realized,
we recast the light neutrino mass matrix, Eq.~(\ref{eq:m-sesa}),
\begin{align*}
  m_{\nu}=-m_D \frac{1}{M_M} m_D^{\rm{T}},
\end{align*}
with $m_D=\lambda \,v$. Here, it is always possible to choose
a basis in which the heavy neutrino mass matrix is diagonal,
$D_M=\text{diag} \left( M_1,M_2,M_3 \right)$. Using an unitary matrix
$U$, one can simultaneously diagonalize the light neutrino mass matrix
\begin{align}
  \label{eq:diagonal-mnu}
  D_m=-\,U^{\dagger}\,m_{\nu}\,U^{\ast},
\end{align}
where $D_m=\text{diag}\left( m_1,m_2,m_3 \right)$.  If one does not
account for the running of neutrino parameters from the electroweak
scale to the see-saw scale~\cite{Babu:1993qv,Antusch:2003kp}, the
matrix $U$ corresponds to the
\emph{Pontecorvo--Maki--Nakagawa--Sakata} (PMNS) matrix. In the basis
where the charged leptons mass matrix is diagonalb and with the help of
an orthogonal matrix $\Omega$, the Dirac neutrino mass matrix can be
written in the so-called \textit{Casas--Ibarra
  parametrization}~\cite{Casas:2001sr},
\begin{align}
  \label{eq:Yukawa-parametrization}
  m_D=U \sqrt{D_m}\,\Omega\, \sqrt{D_M}.
\end{align}
The Dirac neutrino mass matrix is fully described by 18 parameters:
the mixing matrix $U$ contains six parameters (three mixing angles
and three phases), the diagonal matrices $D_m$ and $D_M$ contain three
neutrino masses each, and the orthogonal matrix $\Omega$ is described
by six real (three complex) parameters. It can be written as a product
of three rotational matrices~\cite{DiBari:2005st,Blanchet:2006dq}
\begin{align}
  \label{eq:omega-matrix}
  \Omega \left( \omega_{21},\omega_{31},\omega_{32} \right)=R_{12}
  \left( \omega_{21} \right)\, R_{13} \left( \omega_{31}
  \right)\,R_{23} \left( \omega_{32} \right),
\end{align}
with
\begin{align}
  \label{eq:R12}
  R_{12}\equiv\begin{pmatrix} \pm\sqrt{1-\omega_{21}^2} & -\omega_{21} & 0 \\
    \omega_{21} & \pm\sqrt{1-\omega_{21}^2} & 0 \\ 0 & 0 &
    \pm1 \end{pmatrix},
\end{align}
\begin{align}
  \label{eq:R13}
  R_{13}\equiv\begin{pmatrix} \pm\sqrt{1-\omega_{31}^2} & 0 & -\omega_{31} \\
    0 & \pm 1 & 0 \\ \omega_{31} & 0 &
    \pm\sqrt{1-\omega_{31}^2} \end{pmatrix},
\end{align}
and 
\begin{align}
  \label{eq:R23}
  R_{23}\equiv\begin{pmatrix} \pm1 & 0 & 0 \\
    0 & \pm\sqrt{1-\omega_{32}^2} & -\omega_{32} \\ 0 & \omega_{32} &
    \pm\sqrt{1-\omega_{32}^2} \end{pmatrix}.
\end{align}
In general, one can state that Eq.~(\ref{eq:Yukawa-parametrization})
is divided into two parts: (i) a measurable low-energy part,
containing the PMNS matrix $U$ and the light neutrino masses $D_m$,
and (ii) a high-scale part, consisting of the orthogonal $\Omega$
matrix and the heavy neutrino masses $D_M$, which is not accessible by
current experiments.

The $N_1$-dominated scenario can now be realized assuming the heavy
neutrino mass matrix to be hierarchical, i.e. $M_1\ll M_2 \ll M_3$,
together with a specific choice of the $CP$ asymmetries implemented by
a special form of the $\Omega$
matrix~\cite{DiBari:2005st,Blanchet:2006dq}:
\begin{itemize}
\item For $\Omega=R_{13}$, the $CP$ asymmetry in $N_2$ decays vanishes,
  i.e., $\varepsilon_2=0$, while $\varepsilon_1$ is maximal.
\item For $\Omega=R_{12}$, the $CP$ asymmetry $\varepsilon_1$ is
  suppressed compared to its maximal value, and $\vert \varepsilon_2
  \vert \propto \left( M_1/M_2 \right) \vert \varepsilon_1\vert$ is
  negligible within a strong mass hierarchy~\footnote{If $M_1\sim
    M_2$, both $CP$ asymmetries should be taken into account.}.
\end{itemize} 
In~\cite{Giudice:2003jh,Buchmuller:2004nz,HahnWoernle:2008pq} it has
been shown that the $N_1$-dominated scenario proves to be independent
of the initial conditions in the regime of strong wash-out. However,
in the weak wash-out regime the final asymmetry production depends
sensibly on the initial conditions of the $N_1$ abundance.
Furthermore, thermal leptogenesis sets a lower limit of $M_1 \gtrsim
10^9\GeV$~\cite{Davidson:2002qv} on the mass of the lightest
right-handed neutrino in order to explain successfully the observed
value of the baryon asymmetry of the universe. This bound is
consequently translated into a lower bound on the reheating
temperature after inflation, $T_{RH}\gtrsim 1.5 \times
10^9\GeV$~\cite{Buchmuller:2003gz}. These lower bounds not only have
an issue with the cosmological abundance of
gravitinos~\cite{Moroi:1993mb,Bolz:2000fu,Kawasaki:2004qu,Pradler:2006qh}
within supersymmetric theories, but also cause some specific problems
in grand unified theories based on flavor models. Some of these models
assume a grand unified symmetry between up-quarks and neutrinos. The
neutrino Yukawa couplings are then connected with the up-quark Yukawa
matrices leading to right-handed neutrino masses that are
proportional to the square of the up-quark
masses~\cite{Davidson:2003cq,Branco:2002kt}. Typical values for the
mass of the lightest right-handed state fall in the range
$10^6-10^7\GeV$, see e.g.,~\cite{Babu:1998wi,King:2001uz}, and masses
$M_1\geq 10^9\GeV$ need a specific choice of
parameters~\cite{Raidal:2004vt,Vives:2005ra}. This makes thermal
leptogenesis difficult to reconcile with this class of models.

In order to circumvent these issues, the $N_2$-dominated scenario was
proposed in~\cite{DiBari:2005st}. Indeed, for $\Omega=R_{23}$ one can
have a maximal $CP$ asymmetry $\varepsilon_2$ coming from $N_2$ decays
while $\varepsilon_1$ vanishes. This means that $N_1$ is totally
decoupled from the heavier states while in the $N_2$ decays the heavy
third state, $N_3$, runs in the internal lines of the one-loop
diagrams. Together with a mass hierarchy in the heavy neutrinos,
$N_2$-dominated leptogenesis can be realized if the wash-out from
$N_1$ interactions does not deplete the produced asymmetry. With the
above choice of $\Omega$, the effective neutrino mass of the lightest
right-handed state is fixed to $\tilde{m}_1=m_1$. Thus, for
hierarchical light neutrino masses the $N_1$ interactions can be
forced to be in the weak wash-out regime where $\tilde{m}_1\lesssim
m^{\ast}\approx 10^{-3}\eV$. It is worth noticing that, if the $N_1$
interactions are in the weak wash-out regime, i.e.,
$\tilde{m}_1<m^{\ast}$, then the $N_2$ interactions are constrained to
be in the strong wash-out regime, i.e., $\tilde{m}_2>m^{\ast}$. This
is due to the orthogonality of the $\Omega$
matrix~\cite{DiBari:2005st}. Therefore, $N_2$-dominated leptogenesis
is independent of the initial conditions on $N_2$.  In order to
achieve a large enough $CP$ asymmetry, the lower bound on the mass
$M_1$ can be directly translated into a bound on the mass $M_2$,
whereas the lower bound on the reheating temperature remains
unchanged. The lower bound on $M_1$ is then evaded and $M_1$ can be
arbitrarily light compatibly with the see-saw condition $M \gg m_{D}$.
However, allowing for small complex rotations $R_{12}$ and $R_{13}$,
both of these bounds become increasingly more stringent leading to a
point beyond which the $N_2$-dominated scenario is not viable anymore.

\subsection*{Note on flavor}
\label{sec:NoteFlavor}

Considering flavor effects beyond the alignment case may substantially
change the parameter ranges in which the scenarios discussed above are
valid. When the flavor structure of the lepton asymmetry and the
wash-out is tracked, it is possible to generate a large enough lepton
asymmetry in $N_2$ decays even if $N_1$ interactions are effective. It
has been shown in~\cite{Vives:2005ra} that the $N_1$-induced wash-out
of an asymmetry, established in the lepton state $l_2$ via $N_2$
decays~\footnote{The states $\vert \l_i\rangle=1/
  \sqrt{\left(\lambda^{ \dagger}\lambda \right)_{ii}}
  \,\sum\limits_{\alpha}\lambda_{\alpha i}\vert l_{\alpha}\rangle$ are
  participating in interactions with the heavy states
  $N_i$\cite{Blanchet:2008hg}.}, can be sizably effected by the
projector $P_{1\alpha}=\vert \langle l_1\vert l_{\alpha} \rangle
\vert^2$ that leads to a reduction of the wash-out strength, i.e.,
$K_{1\alpha}< K_{1}$. Another possible modification is that the
coherence of the lepton state $l_2$ is destroyed by potentially fast
occurring $N_1$ inverse decays. This leads to a statistically mixed
state consisting of $l_1$ and the state orthogonal to $l_1$. As a
result, parts of the asymmetry, stored in the state orthogonal to
$l_1$, are protected from $N_1$
wash-out~\cite{Barbieri:1999ma,Strumia:2006qk,Engelhard:2006yg}. This
is only possible when the $N_1$ states decay in the two-flavor regime
at temperatures $10^9\GeV\lesssim T\lesssim 10^{12}\GeV$, where only
the $\tau$ Yukawa interactions are in equilibrium, or in the
unflavored regime at temperatures $T\gtrsim 10^{12}\GeV$. In the fully
flavored regime, however, the full flavor basis is resolved and there
does not exist any direction in flavor space that is protected against
$N_1$ wash-out\cite{Blanchet:2008hg}. The study
of~\cite{Blanchet:2008pw} considered the case $\Omega=R_{12}\left(
  \omega_{21} \right)R_{13}\left( \omega_{31} \right)$ together with
$\omega_{32}=0$ for values $M_1\gtrsim 10^9\GeV$, lying in the
two-flavor regime. It was found that even for small values
$\omega_{21}\leq 0.1$ the contribution to the asymmetry from $N_2$
decays can be naturally the dominant one for $10^{11}\GeV \lesssim M_2
\lesssim 10^{14}\GeV$, when the effects of flavor on the $N_1$-induced
wash-out are taken into account.

Furthermore, in Section~\ref{sec:introduction} we did not account for
the effect of flavor on the $CP$ asymmetry. In
Eq.~(\ref{eq:asy-general}) we wrote the $CP$ asymmetry as a matrix in
flavor space. Indeed, the hierarchy of $CP$ asymmetries, following a
mass hierarchy of particles in the internal lines, does not necessarily
hold when flavored asymmetries are
considered~\cite{Blanchet:2006be}. In contrast to the unflavored
scenario, where $\varepsilon_2$ is suppressed by $M_1/M_2$ compared to
$\varepsilon_1$ when assuming $M_1\ll M_2 \ll 10^{14}\GeV\ll M_3$,
there is the possibility of having a non-negligible asymmetry
$\varepsilon_{2 \alpha}$ that, in turn, eventually extends the range
of the $N_2$-dominated scenario.

\section{Full Boltzmann equations for leptogenesis}
\label{subsec:FullBE}
In this paper we study the wash-out of a lepton asymmetry by means of
the full Boltzmann equations at the mode level including scattering
with the top quark in the alignment case where $K_{i\alpha}=K_{i}$ and
$K_{i\alpha\neq \beta}=0$. This study does not rely on the assumption
of kinetic equilibrium for right-handed neutrinos and will not
restrict the distribution functions to be classical. In order to study
the effects of quantum statistical distribution functions and
scattering with top quarks we specify different scenarios shown in
Table~\ref{tab:scat}.
\begin{table}[t!]  
\centering
\caption{Scenarios in the decays/inverse decays pictures (D1 and D4) and
  scenarios including scattering with the top quark (S1 and S2) 
  and their associated assumptions. 
  Cases D1 and S1 correspond to the conventional
  integrated Boltzmann approach, while Cases D4 and S2 involve 
  solving the full
  set of Boltzmann equations at the mode level. 
  The naming of the different scenarios corresponds
  to~\cite{HahnWoernle:2009qn}.} \vspace{.5em}
\label{tab:scat}
\begin{tabular}[centered]{c c c c}
  \toprule Case & kinetic equilibrium 
  & quantum statistics\\  \midrule
  D1 & Yes & No\\
  D4 & No & Yes\\
  S1 & Yes & No\\
  S2 & No & Yes\\
  \bottomrule
\end{tabular}
\end{table}
The mode equations including scattering processes based on the Yukawa
coupling with the top quark for right-handed neutrinos and the lepton
asymmetry are given by
\begin{align}
  \label{eq:be-rhn-fund-scat}
  \frac{H(M_i)}{z_i}\, \frac{\partial f_{N_i}}{\partial z_i}
  &=C_{D}\left[f_{N_i}\right]+2\,C_{S,s}
  \left[f_{N_i}\right]+4\,C_{S,t}\left[f_{N_i}\right],
  \\[5pt] \label{eq:be-asy-fund-scat} \frac{H(M_i)}{z_i}\,
  \frac{\partial f_{\left(l-\overline{l}\right)_{\alpha}}}{\partial
    z_i}
  &=C_{D}\left[f_{\left(l-\overline{l}\right)_{\alpha}}\right]+2\,C_{S,s}
  \left[f_{\left(l-\overline{l}\right)_{}a}\right]
  +4\,C_{S,t}\left[f_{\left(l-\overline{l}\right)_{\alpha}}\right],
\end{align}
where $z_i=M_i/T$.  The right-hand sides of
Eqs.~(\ref{eq:be-rhn-fund-scat}) and (\ref{eq:be-asy-fund-scat})
contain one collision integral from decays/inverse decays and two
collision integrals from scattering processes coming, respectively,
from scattering in the $s$-channel and in the $t$-channel at tree
level. Here, one factor of 2 stems from contributions from processes
involving anti-particles, and another factor of 2 in the $t$-channel
terms originates from the $u$-channel diagram.

In our analysis we only include tree-level scattering processes that
are of
$\mathcal{O}\left(h^{2}_{t}\lambda^{2}\right)$\cite{HahnWoernle:2009qn}. We
do not consider interactions with gauge bosons, nor include $CP$
violation in $2\rightarrow 2$ or $1(2)\rightarrow 3$ processes, which
are of higher order in the Yukawa couplings. The effect of $CP$
violation from these processes was considered
in~\cite{Abada:2006ea,Pilaftsis:2003gt,Pilaftsis:2005rv,Nardi:2007jp}.
The numerical integration of Eqs.~(\ref{eq:be-rhn-fund-scat})
and~(\ref{eq:be-asy-fund-scat}) over the right-handed neutrino
(lepton) phase space results in the time evolutions of the number
densities $n_{N_i}$ and
$n_{\left(l-\overline{l}\right)_{\alpha}}$. Normalization to the
equilibrium photon number density $n_{\gamma}^{\rm eq}=
(\zeta(3)/\pi^2) g_\gamma T^3$, with $g_\gamma=2$, gives the comoving
abundances $N_{N_i}$ ($N_{\left(l-\overline{l}\right)_{\alpha}}$).
The decay collision integrals for the right-handed neutrinos and the
lepton asymmetry can be evaluated analytically to
be~\cite{Basboll:2006yx,Garayoa:2009my}
\begin{align}
  \label{eq:CI-d-rhn}
  C_{D}\left[f_{N_i}\right]&=\frac{\Gamma_{D_{i}}\,z_i}{\myE_{N}\,
    y_{N}} \,f_{N_{i}}^{\rm{eq}}\, \left( -1 + f_{N_i} + e^{\myE_{N}}
    f_{N_{i}}\right) \, \log \left [\frac{
      \sinh((\myE_{N}-y_{N})/2)}{\sinh((\myE_{N}+y_{N})/2)} \right ],
  \\[5pt] \label{eq:CI-d-asy}
  C_{D}\left[f_{\left(l-\overline{l}\right)_{\alpha}}\right] &=
  -\frac{\Gamma_{D_{i\alpha}}\,z_i}{2y^{2}_{l}}
  \int_{\frac{z_i^{2}-4y^{2}_{l}}{4y_{l}} }^{\infty}
  dy_{N}\,\frac{y_{N}}{\myE_{N}} \left [
    (f_{\Phi}+f_{N_{i}})(f_{\left(l-{\overline{l}}\right)_{\alpha}} +
    \varepsilon_{i\alpha}\, F^{+}_{\alpha}) - 2
    \varepsilon_{i\alpha}\,f_{N_i}\,(1+f_{\Phi}) \right ] \nonumber\\
  &+ \mathcal{O} \left( \lambda_{\alpha i}^2\,\lambda_{\beta i}^2
    \times f_{\left(l-\overline{l}\right)_{\beta}}\right),
\end{align}
where $F^{+}_{\alpha}=f_{l_{\alpha}}+f_{\overline{l}_{\alpha}}$,
$\;\myE_j=E_j/T$, $y_j=\vert \mathbf{p}_j\vert/T$,~\footnote{For
  better clearness we drop the generation index for the heavy
  neutrinos in $\myE_N$ and $y_N$ and the flavor index for the leptons
  in $y_{l}$.}  and $f_{\Phi}$ denotes the distribution function of
the Higgs particle. Up to subleading corrections, that are of higher
order in the Yukawa couplings, the collision integrals in the
different flavors decouple and the corresponding Boltzmann equations
can be solved in analogy to the single flavor
case.\footnote{Reference~\cite{Garayoa:2009my} includes terms in
  addition to Eq.~(\ref{eq:CI-d-asy}) in order to avoid asymmetry
  production in thermal equilibrium.  However, the same analysis also
  shows that the quantitative difference between this and our approach
  is negligible.}  For SM particles we assume Fermi--Dirac
distributions for leptons and a Bose--Einstein distribution for the
Higgs particle, since gauge interactions lead to very fast
equilibration rates\cite{Kolb:1979qa}.

The scattering collision integrals are nine-dimensional and cannot be
evaluated analytically. However, they can be reduced analytically to
two-dimensional integrals given by~\cite{HahnWoernle:2009qn}
\begin{align}
  \label{eq:CI-general-2dim-N}
  C_{S,(s,t)} \left[ f_{N_i} \right] &= \sum_\mu
  \frac{3\,T}{2^{6}\pi^{3}\,\myE_{N}\, y_{N}}\,\frac{h_t^{2}(T)
    \,M_i\,\tilde{m}_{1}}{v^{2}} \\[5pt]
  &\hphantom{\sum_{\mu}}\times
  \int_{w(\myE_{N,l,q,t})}^{u(\myE_{N,l,q,t})}\,
  d\myE_{l}\,\int_{l(\myE_{N,l,q,t})}^{k(\myE_{N,l,t,q})}\,
  d\myE_{(t,q)}\;\Lambda_{(s,t)}^{(N)}\;I_{(s,t)}^{(\mu)},
\end{align}
for the right-handed neutrino and 
\begin{align}
  \label{eq:CI-general-2dim-asy}
  C_{S,(s,t)} \left[ f_{\left(l-\overline{l}\right)_{\alpha}} \right]
  &= \sum_\mu \frac{3\,T}{2^{6}\pi^{3}\,\myE_l^2}\,\frac{h_t^{2}(T)
    \,M_i\,\tilde{m}_{1}}{v^{2}} \nonumber\\
  &\hphantom{\sum_{\mu}} \times
  \int_{w(\myE_{N,l,q,t})}^{u(\myE_{N,l,q,t})}\,
  d\myE_N\,\int_{l(\myE_{N,l,q,t})}^{k(\myE_{N,l,t,q})}\,
  d\myE_{(t,q)}\;\Lambda_{(s,t)}^{(l-\overline{l})_{\alpha}}\;I_{(s,t)}^{(\mu)},
\end{align}
for the asymmetry. Here, $h_{t}^{2}(T)$ is the top Yukawa coupling
evaluated at the leptogenesis scale $z_i$ and the functions
$\Lambda_{(s,t)}^N$ and $\Lambda_{(s,t)}^{(l-\overline{l})_{\alpha}}$
denote the phase space factors for the right-handed neutrino (lepton)
scatterings in the $s$- and $t$-channel, respectively. The integration
limits $u,w,k$, and $l$ depend on the energies of the particles
involved in the interactions and give the distinct integration ranges
for each integrand in the sums of Eqs.~(\ref{eq:CI-general-2dim-N})
and ~(\ref{eq:CI-general-2dim-asy}). The integrands consist of the
phase space factors and the analytical functions $I_{(s,t)}^{\mu}$,
that encode the matrix element and the result of the angle
integration. Both depend on the energies of the particles involved in
the process. For the $s$-channel diagram the sum contains six terms
for the right-handed neutrino and the lepton asymmetry and for the
$t$-channel diagram one counts four terms each. The explicit
expressions of the single collision integrals are listed in the
Appendices~\ref{subsec:full-s-channel} and~\ref{subsec:full-t-channel}
and the rather lengthy derivation can be found in the appendices
of~\cite{HahnWoernle:2009qn,Hahn-Woernlephd}.

Inserting Eqs.~(\ref{eq:CI-general-2dim-N}) and
~(\ref{eq:CI-general-2dim-asy}) in the Boltzmann
equations~(\ref{eq:be-rhn-fund-scat}) and ~(\ref{eq:be-asy-fund-scat})
yields the complete set of differential equations to be integrated
numerically. Once the Boltzmann equations for the distribution
functions are solved, one can perform the integration over the
right-handed neutrino (lepton) phase space to obtain the number
densities and, in turn, the efficiency factors
\begin{align}
  \label{eq:kf-nl}
  \kappa_{i\alpha}=\frac{4}{3}\frac{\varepsilon^{-1}_{i\alpha}}
  {\alpha_{\mathrm{sph}}-1}
  \,N_{\left(l-\overline{l}\right)_{\alpha}}.
\end{align}
A value of $\sum_{\alpha}N_{\left(l-\overline{l}\right)_{\alpha}}
  =3.14\times 10^{-8}$ is needed in order to explain the value
  $\eta_B^{\rm{CMB}}=(6.225\pm0.17)\times 10^{-10}$, measured in
  Cosmic Microwave Background (CMB) observations by the WMAP
  satellite~\cite{Dunkley:2008ie}.

With the restriction on classical equilibrium distribution functions
of the Maxwell--Boltzmann type and assuming kinetic equilibrium holds
for the heavy right-handed neutrinos, Eqs.~(\ref{eq:be-rhn-fund-scat})
and~(\ref{eq:be-asy-fund-scat}) can be integrated over the
right-handed neutrino (lepton) phase space to recast the Boltzmann
equations in an integrated form~\cite{Buchmuller:2004nz},
\begin{eqnarray}
  \label{eq:BE-Nn-s-1} \frac{d N_{N_i}}{d
    z}&=-\,\left(D_i+S_i\right)\, \left(N_{N_i}-N_{N_i}^{\rm eq}\right),\\
  \label{eq:BE-Na-s-1} \frac{d N_{\left(l-\overline{l}\right)\alpha}}{d z} &= -
  W_{i\alpha}\,N_{\left(l-\overline{l}\right)_{\alpha}}
  +\varepsilon_{i\alpha}\,D_i\left(N_{N_i}-N_{N_i}^{\rm eq}\right),
\end{eqnarray}
where $N_j=n_j/n_{\gamma}^{\mathrm{eq}}$. The quantities $D_i$ ($S_i$)
account for the production of right-handed neutrinos from inverse
decays (scatterings), and the wash-out rate $W_{i\alpha}$ contains as
well contributions from both, scatterings ($W^{S}_{\alpha}$) and
inverse decays $W^{ID}_{i\alpha}=P^0_{i\alpha}\,W^{ID}_{i}$. The
explicit expressions for $D_i$, $S_i$, $W_{i\alpha}$, and
$N_{N_i}^{\mathrm{eq}}$ can be found in
Appendix~\ref{subsec:reacrates}.

\section{Mode
  equations in \boldmath{$N_2$}-dominated leptogenesis}
\label{subsec:ful-n2domination}

\begin{figure}[t]
  \centering
  \includegraphics[width=1.0\textwidth]{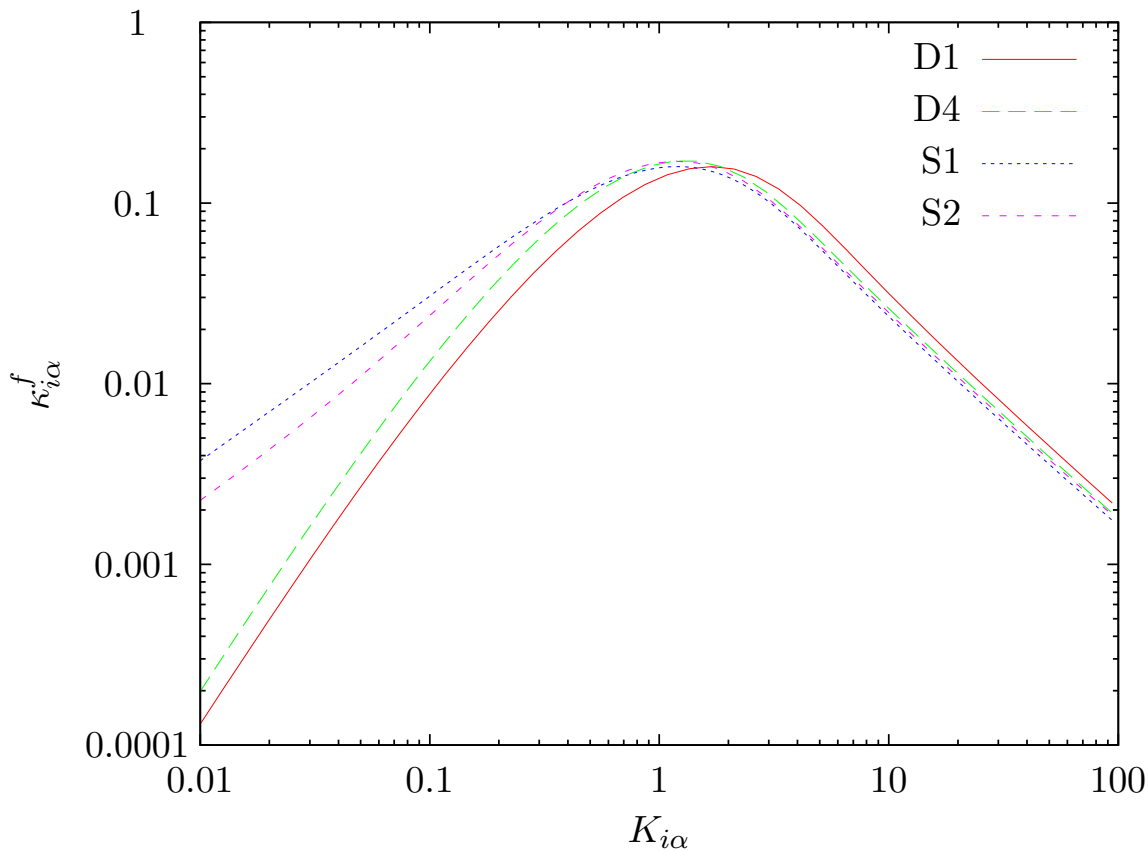}
  \caption{The final efficiency factors $\kappa^{f}_{i\alpha}$ at the
    production stage for a vanishing initial $N_i$ abundance in
    dependence of the interaction parameter $K_{i\alpha}$. Shown
    are the two cases including scattering processes S1 (dotted/blue)
    and S2 (short dashed/magenta), and two scenarios D1 (solid/red)
    and D4 (long dashed/green) within the decay--inverse decay only
    framework~\cite{HahnWoernle:2009qn}.}
  \label{fig:k_full-scat}
\end{figure}

The general effects of the full spectral distributions of right handed
neutrinos on the production of a lepton asymmetry in thermal
leptogenesis have been discussed extensively
in~\cite{HahnWoernle:2009qn}, where the final efficiency factors for
the scenarios, presented in Table~\ref{tab:scat}, have been
calculated. The results can be summarized in
Figure~\ref{fig:k_full-scat}, where the final efficiency factors as a
function of $K_{i\alpha}$ are shown for the integrated approach
(Cases D1 and S1) and the complete mode treatment (Cases D4 and S2),
both including and excluding scattering.  For Cases S1 and S2 which
include scattering, we note that their difference is rather large in
the weak wash-out regime ($K_{i\alpha}<1$),with the integrated
approach overestimating $\kappa^{f}_{i\alpha}$ by up to a factor $\sim
1.5$ at $K_{i\alpha}\sim 0.01$ compared to solving the complete
mode equations. But this difference decreases as we increase
$K_{i\alpha}$.  At $K_{i\alpha} \gtrsim3$ the integrated
approach underestimates $\kappa^{f}_{i\alpha}$ by less than $\sim
10$\%.  It is also interesting to note that the relative contribution
of scattering processes to the final efficiency factor is smaller in
the complete mode calculation than in the integrated approach.  In the
weak wash-out regime, including scattering enhances the final
efficiency factor from decays and inverse decays by up to a factor of
$\sim 30$ in the integrated scenario.  In the complete mode
calculation, however, the enhancement is only a factor of $\sim 15$.
Similarly, in the strong wash-out regime, scattering reduces
$\kappa^{f}_{i\alpha}$ by up to $20$\% in the integrated picture,
compared to below $10$\% in the complete treatment.

In this section, however, we want to discuss the effects the complete
set of mode equations, cf. Eqs.~(\ref{eq:be-rhn-fund-scat})
and~(\ref{eq:be-asy-fund-scat}), have in the $N_2$-dominated scenario.

Here, the wash-out due to $N_1$
interactions is (i) enhanced using mode-equations with decays/inverse
decays alone and (ii) enhanced including scatterings with the top
quark.
\begin{figure}[h!]
  \centering
  \includegraphics[width=1.0\textwidth]{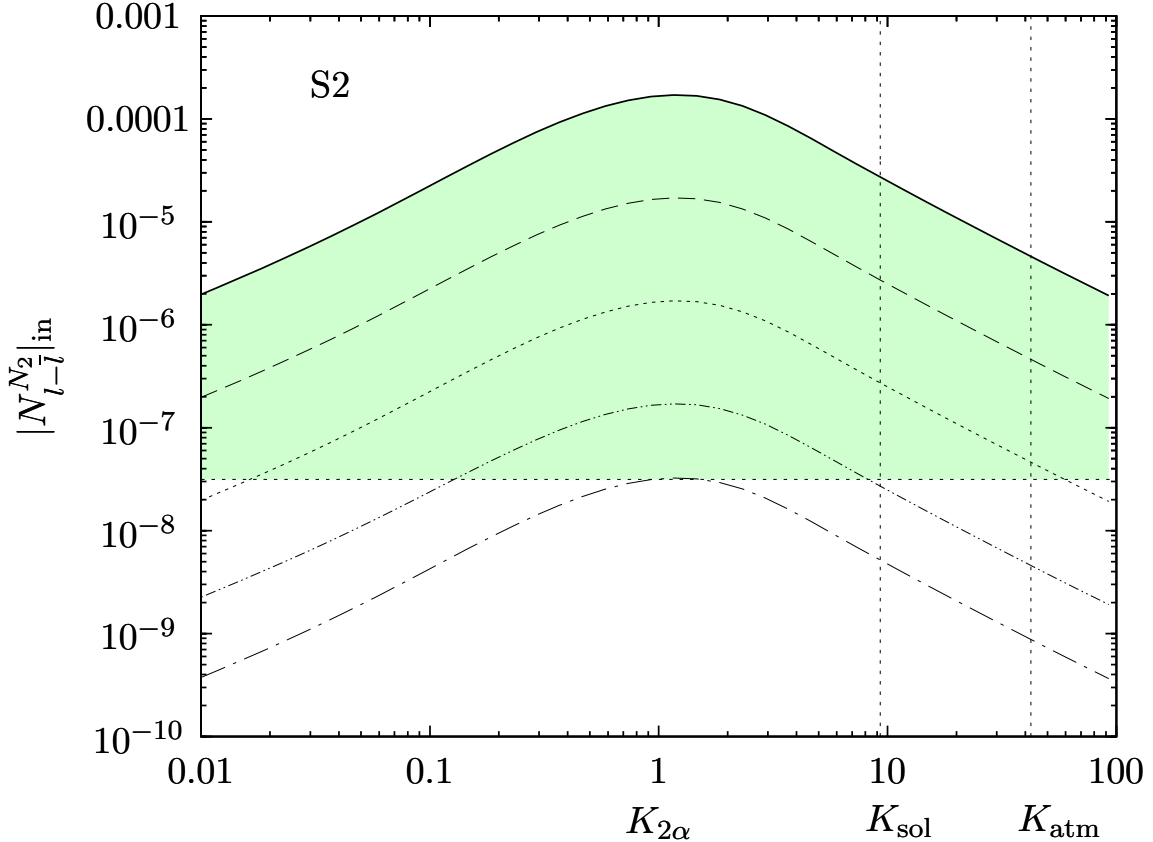} 
  \caption[Lepton asymmetry generated in $N_2$ decays]{Lepton
    asymmetry generated in $N_2$ decays for
    $\varepsilon_{2\alpha}=10^{-6} \left(
      M_2/10^{10} \GeV \right)$ and different values of $M_2$ in the
    Case S2 for a vanishing initial $N_2$ abundance: (i)
    $M_2=10^{13}\GeV$, solid line, (ii) $M_2=10^{12}\GeV$, long-dashed
    line, (iii) $M_2=10^{11}\GeV$, dashed line, (iv)
    $M_2=10^{10}\GeV$, point-point-dashed line, and (v)
    $M_2=1.9\times10^{9}\GeV$, point-dashed line. Within the
    considered scenario, the shaded region indicates where the
    produced baryon asymmetry exceeds $\eta_B^{\rm{CMB}}$. The allowed
    values of $\vert N^{N_2}_{l-\overline{l}}\vert_{\rm in}$ will be
    taken as initial values to calculate the wash-out induced by
    $N_1$-interactions. The differences induced when considering the
    Cases D1, D4 and S1 can directly be read-off from
    Figure~\ref{fig:k_full-scat}.}
  \label{fig:asy_scatN2}
\end{figure}
We will not study $N_2$-dominated leptogenesis with a specific
form for the $\Omega$ matrix but rather consider values of
$K_{1\alpha}$ and $K_{2\alpha}$ going from the weak wash-out
regime ($K_{i\alpha}=0.01$) up to the strong wash-out regime
($K_{i\alpha}=100)$ without the restrictions that are usually
imposed when choosing a specific $\Omega$ matrix constellation. For
example one usually has $K_{2\alpha}\gtrsim K_{\mathrm{sol}}\sim 9$
for $\Omega_{23}$\cite{DiBari:2005st,Blanchet:2008pw}.  In doing so we
can study the effect of the enhanced wash-out due to scattering and
the use of mode equations on the $N_1$-induced wash-out and restrict
the general available parameter space for $N_2$-dominated leptogenesis
in terms of $M_{1,2}$ and $K_{(1,2)\alpha}$. A similar study has
been performed in\cite{Garayoa:2009my} where, however, wash-out
effects have been considered in the decay-inverse decay only picture.

The lepton asymmetry that can be generated in $N_2$ decays is shown in
Figure~\ref{fig:asy_scatN2} in dependence of the decay parameter
$K_{2\alpha}$ for values of $M_2$ varying between $1.9 \times
10^{9}\GeV$ and $10^{13}\GeV$. The shaded area marks the region where
the final asymmetry exceeds the value $\eta_B^{\rm{CMB}}$, measured in
the observation of the Cosmic Microwave Background. For $M_2\lesssim 2
\times 10^9\GeV$ the $CP$ asymmetry generated in $N_2$ decays is too
small to account for the observed value of the matter--antimatter
asymmetry.~\footnote{This bound on $M_2$ is more severe than the value
  given in Section~\ref{subsec:n2-realization} since for a vanishing
  initial abundance $\kappa^{f}_{i\alpha}<1$.} We assumed the maximal
value of the $CP$ asymmetry
$\varepsilon^{\textrm{max}}_{2\alpha}\approx 10^{-6} \left(
  M_2/10^{10} \GeV \right)$ to be realized. In specific scenarios,
however, the $CP$ asymmetry $\varepsilon_{2\alpha}$ will be smaller
and the corresponding lower bound will be at $M_2\gtrsim
10^{10}\GeV$\cite{DiBari:2005st}.  On the other hand, for $M_2\gtrsim
10^{13}\GeV$ one has to account for $\Delta L=2$ violating scattering
processes with the $N_2$ in the $s$- and $t$-channel. Being an
additional contribution to the wash-out, these processes tend to
reduce the final amount of asymmetry~\cite{Buchmuller:2002rq}. The two
vertical lines in Figure~\ref{fig:asy_scatN2} correspond to the solar
and atmospheric neutrino scales, respectively. Considering the $N_2$
interactions to be in the strong wash-out regime in the window
preferred by neutrino oscillation data demands $M_2\gtrsim
10^{11}\GeV$ to explain the observed value of the baryon asymmetry.
\begin{figure}[h!]
  \centering
  \includegraphics[width=0.9\textwidth]{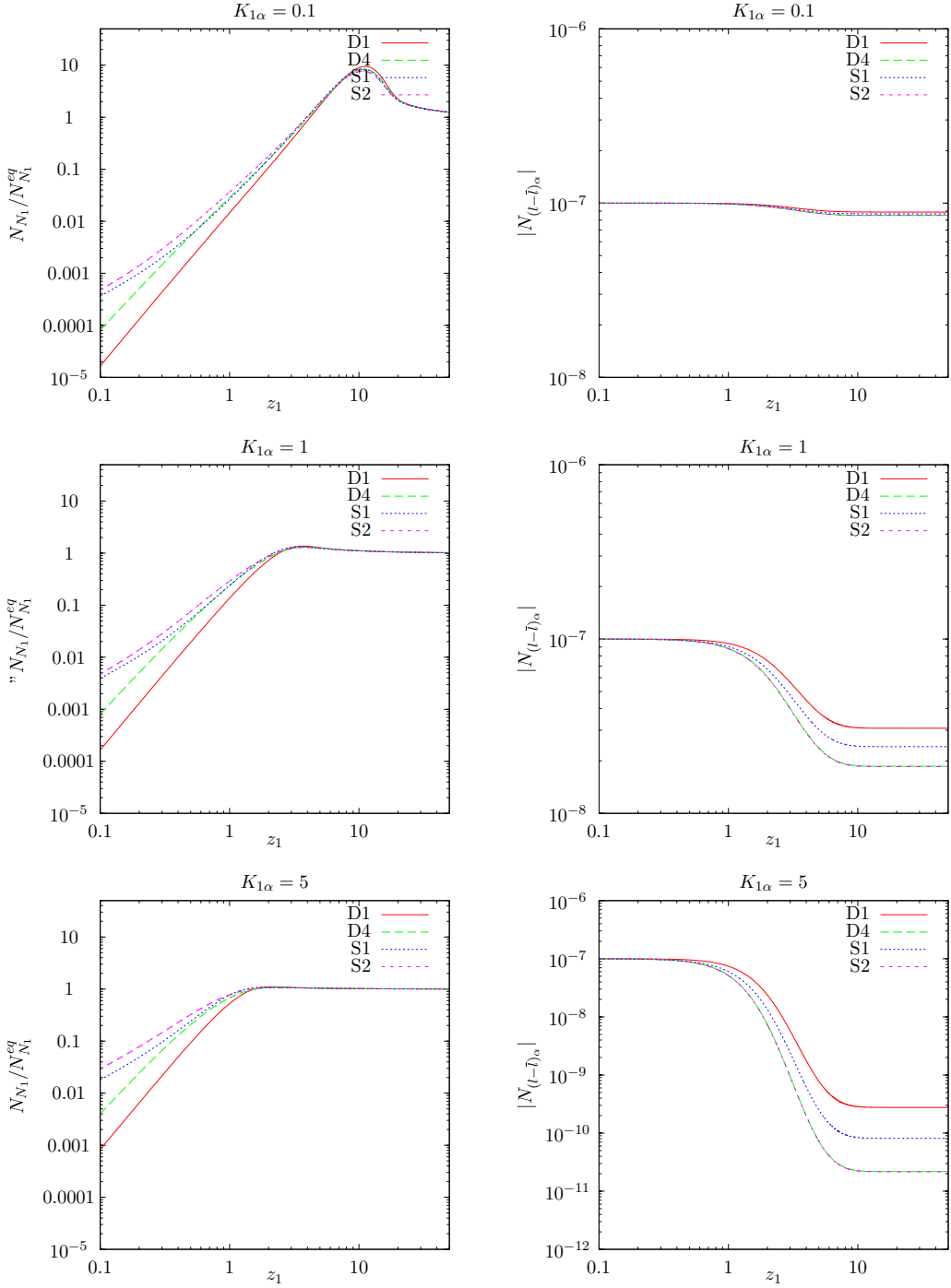} 
  \caption[Time evolution of the right-handed neutrino abundance and
  the lepton asymmetry in $N_2$-dominated leptogenesis]{Time evolution
    of the absolute value of the normalized right-handed neutrino
    number density $N_{N_1}/N_{N_1}^{\rm{eq}}$ and the lepton
    asymmetry $\vert N_{\left(l-\overline{l}\right)_{\alpha}} \vert$
    for three different coupling strengths $K_{1\alpha}$ in the
    $N_2$-dominated scenario for an initial asymmetry $\vert N_{
      l-\overline{l}}^{N_2}\vert_{\rm{in}}=10^{-7}$, $M_1=10^7\GeV$
    and a vanishing initial $N_1$ abundance. The differences between
    the Cases D4 and S2 are smaller than 2 \%.}
  \label{fig:NBL_scatN2}
\end{figure}

The lepton asymmetry generated in $N_2$ decays is altered by the
subsequent wash-out due to interactions of the lightest right-handed
neutrino $N_1$.  According to the considerations
of~\cite{Garayoa:2009my}, where $N_2$-dominated leptogenesis has been
addressed by means of mode equations within the decay--inverse decay
only scenario (Case D4), we choose the following initial conditions at
$z_{1}=M_{1}/T$ to calculate the effect of wash-out on an initially
produced asymmetry: (i) We take $\vert
N_{l-\overline{l}}^{N_{2}}\vert_{\rm{in}}=10^{-7}$ as initial value of
the lepton asymmetry generated in $N_2$ decays, (ii) assume a zero
initial $N_1$ abundance, and (iii) set
$\varepsilon_{1\alpha}\approx 0$. The third condition can be
achieved by supposing a small value of the $N_1$ mass. Anyway, small
lepton asymmetries stemming from different generations add up linearly
and an additional asymmetry $\varepsilon_{1\alpha}$ would not
modify our considerations on the $N_1$-induced wash-out effects. We
choose $M_1=10^7\GeV$ in the numerical implementation in order to fix
the evolution of the top Yukawa coupling.

Figure~\ref{fig:NBL_scatN2} shows on the left panel the time evolution
of the normalized $N_1$ number density for the Cases D1, D4, S1, and
S2 in dependence of $z_{1}$ and $M_1=10^7\GeV$. On the right panel the
time evolution of the lepton asymmetry during $N_1$ wash-out is shown
for the same scenarios. Thus, in addition to the discussion
in~\cite{Garayoa:2009my}, we include Cases S1 and S2 here. In the weak
wash-out regime ($K_{1\alpha}=0.1$), the asymmetry is only slightly
reduced in the Cases D4, S1, and S2 compared to the integrated
approach in the decay-inverse decay only scenario, Case D1. The net
wash-out of the initial asymmetry is less than 10\%. However, already
at $K_{1\alpha}\sim 1$ the strength of the wash-out in the
different scenarios becomes distinguishable. At $z_1\sim 1$ wash-out
becomes effective and is strongest in Case S2 where the complete set
of Boltzmann equations including scattering with the top quark is
considered. However, the net reduction of the initial asymmetry is
still less than one order of magnitude. The difference in the lepton
asymmetry between Case D4 and Case S2, both using mode equations, is
very small (below 2 \%) and cannot be see in
Figure~\ref{fig:NBL_scatN2} .~\footnote{An error in the numerical
  calculation for the Case D4, appearing in an earlier
  publication~\cite{Hahn-Woernlephd}, that leads to larger differences
  between the Cases D4 and S2, has been corrected in the present
  paper.} Comparing Case S2 with Case S1, we see that the influence of
the additional wash-out factor $f_{N_{1}}$, present in the mode
equation, cf. Eq.~(\ref{eq:CI-d-asy}), is larger than the wash-out due
to scatterings in the integrated approach that is present in Case
S1. In the strong wash-out regime, for $K_{1\alpha}=5$, the initial
asymmetry is depleted by up to two orders of magnitude, with the
strongest wash-out again in Case S2. Considering the momentum
integrated scenarios, the reduction in Case S1 is two orders of
magnitude larger than in Case D1. Concerning the scenarios in which
mode equations are used, the contribution of the scatterings amounts
again less than 2 \% as can be seen by comparing Case S2 with Case
D4.  In general, it can be stated that the wash-out is enhanced more
by the use of mode equations (due to an additional wash-out factor
$f_{N_{1}}$) than by the inclusion of scattering processes that become
important only for values $K_{1\alpha}\geq 1$ in the integrated
approach (Case S1).

\begin{figure}[t!]
  \centering
  \includegraphics[width=1.0\textwidth]{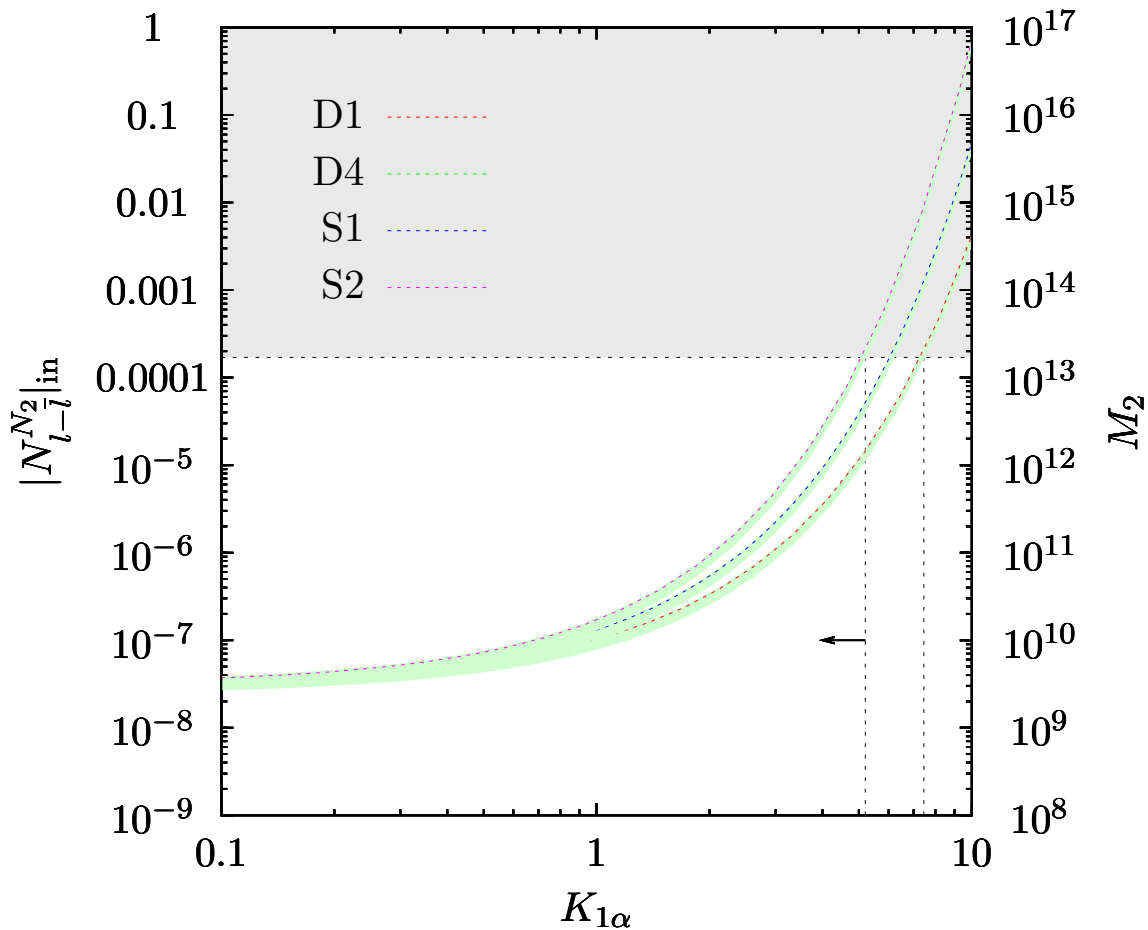}
  \caption[Initial amount of asymmetry needed for successful
  $N_2$-dominated leptogenesis]{Amount of initial asymmetry that has
    to be generated in $N_2$ decays in order to survive the subsequent
    wash-out by $N_1$ interactions for the Cases D1, D4, S1 and
    S2. The mass of the lightest right-handed neutrino was set to
    $M_1=10^7\GeV$ and the $CP$ asymmetry generated in $N_1$ decays
    was set $\varepsilon_{1\alpha}=0$. The mass $M_2^{S2}$, that is
    needed to produce enough initial asymmetry in $N_2$ decays when
    $K_{2\alpha}\approx 1$, is
    shown on the right vertical axis for the Case S2. The asymmetry in
    the gray shaded region cannot be generated in $N_2$ decays and the
    area right of the arrow is excluded due to $N_1$ wash-out.}
  \label{fig:NBL-N2-ini}
\end{figure}
Figure~\ref{fig:NBL-N2-ini} shows the amount of initial asymmetry
generated in $N_2$ decays that is needed to account for the observed
value of the baryon asymmetry after wash-out due to $N_1$ interactions
in dependence of $K_{1\alpha}$ for the Cases D1, D4, S1 and S2. For
the Case S2, the mass of the next-to-lightest neutrino $M_2^{S2}$,
that is needed to produce enough initial asymmetry when
$K_{2\alpha}\approx 1$, is shown on the right vertical axis. The
green colored regions correspond to the value of the baryon asymmetry
deduced in Big Bang Nucleosynthesis (BBN) within 95\% confidence
level~\cite{Yao:2006px}, and the dashed-lines within this regions
represent the value of the baryon asymmetry deduced from CMB
measurements. When $N_1$ interactions fall into the weak-wash-out
regime ($K_{1\alpha}<1$), almost all of the initial generated
asymmetry survives giving \mbox{$\vert
  N_{l-\overline{l}}^{N_2}\vert_{\rm{in}}\approx 3 \times 10^{-8}$},
i.e., the same limit as can be seen in Figure~\ref{fig:asy_scatN2} and
the different Cases D1, D4, S1 and S2 cannot be
distinguished. However, increasing the strength of the $N_1$
interactions, wash-out becomes more and more effective and for
$K_{1\alpha}\gtrsim1$ the differences between the Cases D1, D4 and
S2 become visible. Again, considering the treatment with mode
equations (Cases D4 and S2), the contributions of the scatterings are
below \mbox{2 \%} and too small to be visible in
Figure~\ref{fig:NBL-N2-ini}. For \mbox{$K_{1\alpha}=10$} an initial
value $\vert N_{l-\overline{l}}^{N_2}\vert_{\rm{in}}\approx 1$ is
needed to account for the observed value of the baryon
asymmetry. Though, as can be seen in Figure~\ref{fig:asy_scatN2},
values of the asymmetry lying in the shaded area above the horizontal
dashed line at $\vert N_{l-\overline{l}}^{N_2}\vert_{\rm{in}}=1.7
\times 10^{-4}$ cannot be generated in $N_2$ decays. Therefore, for
the $N_2$-dominated scenario to be successful, the $N_1$ interactions
are restricted to $K_{1\alpha}\lesssim 5$ for the Cases S2 and
D4. When considering the integrated Case D1, this bound can be relaxed
to $K_{1\alpha}\lesssim 7$. Considering Case S2, the upper bound on
the initial asymmetry corresponds to values $M_2^{S2}=10^{13}\GeV$ and
$K_{2\alpha}\approx 1$ as can be seen on the right vertical axis
of Figure~\ref{fig:NBL-N2-ini}. For values of $K_{2\alpha}$ lying
in the strong wash-out regime, that is preferred by neutrino
oscillation data, the generated asymmetry is roughly one order of
magnitude smaller, leading to $K_{1\alpha}\lesssim 3$ in Case S2
and $K_{1\alpha}\lesssim 4$ for Case D1. When choosing conservative
values, $M_2\approx 10^{11} \GeV$ and $\kappa^{f}_{2\alpha}\approx 10^{-2}$,
the scenarios are forced to $K_{1\alpha}\lesssim 2$ for Case S2 and
$K_{1\alpha}\lesssim 3$ for D1.~\footnote{Using typical
  assumptions, the study in~\cite{Garayoa:2009my} found a limit
  $K_{1\alpha}<3$.} This corresponds to values of $K_{1\alpha}$
typically needed in $SO(10)$ inspired GUTs where flavor effects are
included. For the limiting scenario, $M_2\approx 2\times 10^9\GeV$,
the $N_1$ interactions strength $K_{1\alpha}$ has to
vanish~\cite{DiBari:2008mp} in all considered cases.

\section{Conclusions}\label{sec:conclusions}

In this paper we studied leptogenesis in an alternative scenario where
the lepton asymmetry is created in the decays of the next-to-lightest
right-handed neutrino state $N_2$. Here, the additional wash-out
present in the complete set of mode equations leads to a more
efficient depletion of the lepton asymmetry in interactions of the
lightest right-handed neutrino $N_1$; these interactions follow the
asymmetry generation in the decays of the heavier state. In order to
account for the observed value of the matter-antimatter asymmetry, the
possible values of the decay parameters $K_{1\alpha}$ and
$K_{2\alpha}$ of the two right-handed states can be restricted. From
the maximal amount of asymmetry that is achievable in $N_2$ decays,
the decay parameter of the lightest right-handed neutrino is forced to
$K_{1\alpha} \lesssim 5$ when accounting for the spectral distribution
of right-handed neutrinos. The effects of scattering process on the
$N_1$-induced wash-out are smaller than 2 \%.  Furthermore, demanding
the decay parameter $K_{2\alpha}$ to be in the strong wash-out regime
favored by neutrino oscillation data, where the asymmetry generation
is independent of the initial conditions on $N_2$, sets the more
stringent limit $K_{1\alpha}\lesssim 2$.
  
\subsection*{Note added}
\label{sec:NoteAdded}
After the present paper was submitted, a new analysis of leptogenesis
in the $N_2$-dominated scenario appeared~\cite{Antusch:2010ms}, in
which the effects of flavor on the $CP$ asymmetries in $N_2$-dominated
leptogenesis are studied in detail.

\acknowledgments

We thank Steve Blanchet and Michael Pl\"umacher for useful
discussions and comments.

\appendix
\section*{Appendices}

\section{\boldmath{$s$}-channel
  collision integrals}
\label{subsec:full-s-channel}

\subsection*{Right-handed neutrino}\label{sec:right-hand-neutr}
Using energy conservation, and Fermi--Dirac statistics for the leptons
and quarks, the phase space factor reads
\begin{equation}
  \label{eq:lam-s-n-2}
  \Lambda_{s}^{(N_i)}\left(f_{N_i},f_{l_{\alpha}},f_{t},f_{q}\right)=-\,\frac{\,e^{\myE_{l}+\myE_{t}}\;
    \left(-1+f_{N_i}+e^{\myE_{N}}\,f_{N_i}\right)}
  {\left(1+e^{\myE_{l}}\right)\,\left(1+e^{\myE_{t}}\right)\,
    \left(e^{\myE_{l}+\myE_{N}}+e^{\myE_{t}} \right)}.
\end{equation}
The $s$-channel contribution to Eq.~(\ref{eq:CI-general-2dim-N}) is
then given by,
\begin{equation}
 C_{S,s}[f_{N_i}] =
 C_{s}^{(1)}+C_{s}^{(2)}+C_{s}^{(3)}+C_{s}^{(4)}+C_{s}^{(5)}+C_{s}^{(6)},
\end{equation}
The single collision integrals must be evaluated numerically and then
summed to give $C_{S,s}[f_{N_{i}}]$. The integrals
$C_{S,s}^{(1,\ldots,6)}$ are as follows:

\begin{enumerate}
\item First integral (with $\tilde{q}\equiv q/T$):

  \begin{align}
    \label{eq:Cs1}
    C_{S,s}^{(1)}&=\frac{3\,T}{2^{6}\pi^{3}\,\myE_{N}\,
      y_{N}}\,\frac{h_t^{2} \,M_i,\tilde{m}_i}{v^{2}}
    \int_{y_{N}}^{\infty}\,
    d\myE_{l}\,\int_{0}^{\myE_{l}+\myE_{N}}\,d\myE_{t}\;\Lambda_{s}^{(N)}\;I_{s}^{(1)},
    \\ \nonumber
    I_{s}^{(1)}&=\int_{\myE_{l}-y_{N}}^{\myE_{l}+y_{N}}\,d \tilde{q}\,
    \frac{\left(\myE_{N}+\myE_{l}\right)^{2}-z^{2}_i-\tilde{q}^{2}
    }{\left(\myE_{N}+\myE_{l}\right)^{2}
      -\tilde{q}^{2}}  \\
    \label{eq:Is1} &= \frac{4\,y_{N}\,\left(\myE_{l}+\myE_{N}\right) +
      z^{2}_i\, \log \left[\frac{\left(\myE_{N}-y_{N}\right)\,
          \left(2\myE_{l}+\myE_{N}-y_{N}\right)}
        {\left(\myE_{N}+y_{N}\right)\,\left(2\myE_{l}+\myE_{N}+y_{N}\right)}
      \right] }{2\left(\myE_{l}+\myE_{N}\right)}.
  \end{align}

\item Second integral:
  
  \begin{align}
    \label{eq:Cs1-b}
    C_{S,s}^{(2)}&=\frac{3\,T}{2^{6}\pi^{3}\,\myE_{N}\,
      y_{N}}\,\frac{h_t^{2} \,M_i\,\tilde{m}_i}{v^{2}}
    \int_{0}^{y_{N}}\,
    d\myE_{l}\,\int_{0}^{\myE_{l}+\myE_{N}}\,d\myE_{t}\;\Lambda_{s}^{(N)}\;I_{s}^{(2)},
    \\ \nonumber
    I_{s}^{(2)}&=\int_{y_{N}-\myE_{l}}^{\myE_{l}+y_{N}}\,d \tilde{q}\,
    \frac{\left(\myE_{N}+\myE_{l}\right)^{2}-z^{2}_i-\tilde{q}^{2}
    }{\left(\myE_{N}+\myE_{l}\right)^{2}
      -\tilde{q}^{2}}  \\
    \label{eq:Is1-b}
    &=\frac{4\,\myE_{l}\,\left(\myE_{l}+\myE_{N}\right) + z^{2}_i\, \log
      \left[\frac{\myE_{N}^{2}-y_{N}^{2}}
        {\left(2\myE_{l}+\myE_{N}\right)^{2}-y_{N}^{2}} \right]
    }{2\left(\myE_{l}+\myE_{N}\right)}.
  \end{align}

\item Third integral:

  \begin{align}
    \label{eq:Cs2}
    C_{S,s}^{(3)}&=\frac{3\,T}{2^{6}\pi^{3}\,\myE_{N}\, y_{N}
    }\,\frac{h_t^{2} \,M_i\,\tilde{m}_i}{v^{2}} \int_{y_N}^{\infty}\,
    d\myE_{l}\,\int_{0}^{\frac{1}{2}\,\left(\myE_{N}+y_{N}\right)}\,
    d\myE_{t}\;\Lambda_{s}^{(N)}\;I_{s}^{(3)}, \\ \nonumber
    I_{s}^{(3)}&=-\,\int_{\myE_{l}-y_{N}}^{\myE_{l}+\myE_{N}-2\myE_{t}}\,d
    \tilde{q} \,
    \frac{\left(\myE_{N}+\myE_{l}\right)^{2}-z^{2}_i-\tilde{q}^{2}
    }{\left(\myE_{N}+\myE_{l}\right)^{2} -\tilde{q}^{2}}
    \\
    \label{eq:Is2} &=-\, \frac{2\;\left(\myE_{l}+\myE_{N}\right)\,
      \left(\myE_{N}-2\, \myE_{t}+y_{N}\right) + z^{2}_i\, \log
      \left[\frac{\myE_{t}\, \left( 2\,\myE_{l}+\myE_{N}-y_{N}\right)}
        {\left(\myE_{l}+\myE_{N}-\myE_{t}\right)\,
          \left(\myE_{N}+y_{N}\right)} \right]}{2
      \left(\myE_{l}+\myE_{N}\right)}.
  \end{align}
  
\item Fourth integral:

  \begin{align}
    \label{eq:Cs3}
    C_{S,s}^{(4)}&=\frac{3\,T}{2^{6}\pi^{3}\,\myE_{N}\, y_{N}
    }\,\frac{h_t^{2} \,M_i\,\tilde{m}_i}{v^{2}} \int_{y_N}^{\infty}\,
    d\myE_{l}\,\int_{\frac{1}{2}\,\left(2\,\myE_{l}+\myE_{N}-y_{N}
      \right)}^{\myE_{l}+\myE_{N}}\,d\myE_{t}\;\Lambda_{s}^{(N)}\;I_{s}^{(4)}, \\
    \nonumber
    I_{s}^{(4)}&=-\,\int_{\myE_{l}-y_{N}}^{2\myE_{t}-\myE_{l}-\myE_{N}}\,
    d \tilde{q} \,
    \frac{\left(\myE_{N}+\myE_{l}\right)^{2}-z^{2}_i-\tilde{q}^{2}
    }{\left(\myE_{N}+\myE_{l}\right)^{2} -\tilde{q}^{2}}
    \\
    \label{eq:Is3} &= \frac{ 2\;\left(\myE_{l}+\myE_{N}\right)\,
      \left(2\,\myE_{l}+\myE_{N}-2\, \myE_{t}-y_{N}\right) - z^{2}_i\,
      \log \left[\frac{ \left(\myE_{l}+\myE_{N}-\myE_{t}\right)\,
          \left( 2\,\myE_{l}+\myE_{N}-y_{N}\right)}{\myE_{t}\,
          \left(\myE_{N}+y_{N}\right) } \right]}{2
      \left(\myE_{l}+\myE_{N}\right)}.
 \end{align}
  
\item Fifth integral:

  \begin{align}
    \label{eq:Cs4}
    C_{S,s}^{(5)}&=\frac{3\,T}{2^{6}\pi^{3}\,\myE_{N}\, y_{N}
    }\,\frac{h_t^{2} \,M_i\,\tilde{m}_i}{v^{2}} \int_{0}^{y_N}\,
    d\myE_{l}\,\int_{0}^{\frac{1}{2}\,\left(2\,\myE_{l}+\myE_{N}-y_{N}
      \right)}\,d\myE_{t}\;\Lambda_{s}^{(N)}\;I_{s}^{(5)},\\ \nonumber
    I_{s}^{(5)}&=-\,\int_{y_{N}-\myE_{l}}^{\myE_{l}+\myE_{N}-2\myE_{t}}\,
    d \tilde{q} \,
    \frac{\left(\myE_{N}+\myE_{l}\right)^{2}-z^{2}_i-\tilde{q}^{2}
    }{\left(\myE_{N}+\myE_{l}\right)^{2} -\tilde{q}^{2}}
    \\
    \label{eq:Is4} &=-\, \frac{2\;\left(\myE_{l}+\myE_{N}\right)\,
      \left(2\,\myE_{l}+\myE_{N}-2\, \myE_{t}-y_{N}\right) - z^{2}_i\,
      \log \left[\frac{ \left(\myE_{l}+\myE_{N}-\myE_{t}\right)\,
          \left( 2\,\myE_{l}+\myE_{N}-y_{N}\right)}{\myE_{t}\,
          \left(\myE_{N}+y_{N}\right) }
      \right]}{2\left(\myE_{l}+\myE_{N}\right)}.
  \end{align}

\item Sixth integral:

  \begin{align}
    \label{eq:Cs5}
    C_{S,s}^{(6)}&=\frac{3\,T}{2^{6}\pi^{3}\,\myE_{N}\, y_{N}
    }\,\frac{h_t^{2} \,M_i\,\tilde{m}_i}{v^{2}} \int_{0}^{y_N}\,
    d\myE_{l}\,\int_{\frac{1}{2}\,\left(\myE_{N}+y_{N}\right)}^{\myE_{l}+\myE_{N}}\,
    d\myE_{t}\;\Lambda_{s}^{(N)}\;I_{s}^{(6)},\\
    \nonumber
    I_{s}^{(6)}&=-\,\int_{y_{N}-\myE_{l}}^{2\myE_{t}-\myE_{l}-\myE_{N}}\,
    d \tilde{q} \,
    \frac{\left(\myE_{N}+\myE_{l}\right)^{2}-z^{2}_i-\tilde{q}^{2}
    }{\left(\myE_{N}+\myE_{l}\right)^{2} -\tilde{q}^{2}}
    \\
    \label{eq:Is5} &= \frac{2\;\left(\myE_{l}+\myE_{N}\right)\,
      \left(\myE_{N}-2\, \myE_{t}+y_{N}\right) - z^{2}_i\, \log
      \left[\frac{ \left(\myE_{l}+\myE_{N}-\myE_{t}\right)\, \left(
            \myE_{N}+y_{N}\right)}{\myE_{t}\,\left(2\,\myE_{l}+
            \myE_{N}-y_{N}\right) } \right]}{2
      \left(\myE_{l}+\myE_{N}\right)}.
  \end{align}
  
\end{enumerate}

\subsection*{Lepton asymmetry}
\label{sec:lepton-asymmetry-s}
For the lepton asymmetry the $s$-channel phase space element is given by
\begin{align}
  \label{eq:lam-s-a-2}
  \Lambda_{s}^{\left(l-\overline{l}\right)_{\alpha}}
  \left(f_{\left(l-\overline{l}\right)_{\alpha}},f_{N_i},f_{t},t_{q}\right)=
  -\,f_{\left(l-\overline{l}\right)_{\alpha}}\,
  \frac{e^{\myE_{t}}\,\left(1+\left(e^{\myE_{l}+\myE_{N}}-1\right)\,f_{N_i}\right)}
  {\left(1+e^{\myE_{t}}\right)\,\left(e^{\myE_{l}+\myE_{N}}+e^{\myE_{t}}\right)},
\end{align}
and the $s$-channel contribution to Eq.~(\ref{eq:CI-general-2dim-asy})
can be expressed as
\begin{align}
  \label{eq:C-s-l-C}
  C_{S,s}[f_{\left(l-\overline{l}\right)_{\alpha}}] =
  C_{S,s}^{(1)}+C_{S,s}^{(2)}+C_{S,s}^{(3)}
  +C_{S,s}^{(4)}+C_{S,s}^{(5)}+C_{S,s}^{(6)}
\end{align}
to be integrated numerically over two remaining degrees of freedom.
The explicit integrals in Eq.~(\ref{eq:C-s-l-C}) are:
\begin{enumerate}
\item First integral
  \begin{align}
    \label{eq:Cas1}
    C_{S,s}^{(1)}&= \frac{3\,T}{2^{6}\pi^{3}\,
      \myE^{2}_{l}}\,\frac{h_t^{2} \,M_i\,\tilde{m}_i}{v^{2}}
    \int_{z}^{\sqrt{\myE^2_l+z^2_i}}\, d
    \myE_{N}\,\int_{0}^{\myE_{l}+\myE_{N}}\,d\myE_{t}\;
    \Lambda_{s}^{(l-\overline{l})}\; I_{s}^{(1)},
  \end{align}
  where $I_{s}^{(1)}$ is given by Eq.~(\ref{eq:Is1}).

\item Second integral with $I_{s}^{(2)}$ given by 
  Eq.~(\ref{eq:Is1-b}):
   \begin{align}
    \label{eq:Cas1a}
    C_{S,s}^{(2)}&= \frac{3\,T}{2^{6}\pi^{3}\,
      \myE^{2}_{l}}\,\frac{h_t^{2} \,M_i\,\tilde{m}_i}{v^{2}}
    \int_{\sqrt{\myE_l^2+z^2_i}}^{\infty}\, d
    \myE_{N}\,\int_{0}^{\myE_{l}+\myE_{N}}\,d\myE_{t}\;
    \Lambda_{s}^{(l-\overline{l})}\; I_{s}^{(2)}.
  \end{align}

\item Third integral ($I_s^{(3)}$ given by Eq.~(\ref{eq:Is2})):
  \begin{align}
    \label{eq:Cas2}
    C_{S,s}^{(3)}&= \frac{3\,T}{2^{6}\pi^{3}\,
      \myE^{2}_{l}}\,\frac{h_t^{2} \,M_i\,\tilde{m}_i}{v^{2}}
    \int_{z}^{\sqrt{\myE_l^2+z^2_i}}\, d
    \myE_{N}\,\int_{0}^{\frac{1}{2}\,\left(\myE_{N}+y_{N}\right)}\,
    d\myE_{t}\;\Lambda_{s}^{(l-\overline{l})}\; I_{s}^{(3)}.
  \end{align}

\item Fourth integral ($I_s^{(4)}$ given by Eq.~(\ref{eq:Is3})):
  \begin{align}
    \label{eq:Cas3}
    C_{S,s}^{(4)}&= \frac{3\,T}{2^{6}\pi^{3}\,
      \myE^{2}_{l}}\,\frac{h_t^{2} \,M_i\,\tilde{m}_i}{v^{2}}
    \int_{z}^{\sqrt{\myE_l^2+z^2_i}}\, d \myE_{N}\,
    \int_{\frac{1}{2}\,\left(2\,\myE_{l}+\myE_{N}-y_{N}\right)}^{\myE_{l}+\myE_{N}}\,
    d\myE_{t}\;\Lambda_{s}^{(l-\overline{l})}\; I_{s}^{(4)}.
  \end{align}

\item Fifth integral ($I_s^{(5)}$ given by Eq.~(\ref{eq:Is4})): 
  \begin{align}
    \label{eq:Cas4}
    C_{S,s}^{(5)}&= \frac{3\,T}{2^{6}\pi^{3}\,
      \myE^{2}_{l}}\,\frac{h_t^{2} \,M_i\,\tilde{m}_i}{v^{2}}
    \int_{\sqrt{\myE_l^2 + z^2_i}}^{\infty}\, d
    \myE_{N}\,\int_{0}^{\frac{1}{2}\,\left(2\,\myE_{l}+\
        \myE_{N}-y_{N}\right)}\,d\myE_{t}\;\Lambda_{s}^{(l-\overline{l})}\;
    I_{s}^{(5)}.
  \end{align}

\item Sixth integral ($I_s^{(6)}$ given by Eq.~(\ref{eq:Is5})): 
  \begin{align}
    \label{eq:Cas5}
    C_{S,s}^{(6)}&= \frac{3\,T}{2^{6}\pi^{3}\,
      \myE^{2}_{l}}\,\frac{h_t^{2} \,M\,\tilde{m}_i}{v^{2}}
    \int_{\sqrt{\myE^2_l + z^2_i}}^{\infty}\, d
    \myE_{N}\,\int_{\frac{1}{2}\,\left(\myE_{N}+y_{N}\right)}
    ^{\myE_{l}+\myE_{N}}\,d\myE_{t}\;\Lambda_{s}^{(l-\overline{l})}\;
    I_{s}^{(6)}.
  \end{align}

\end{enumerate}
\section{\boldmath{$t$}-channel
  collision integrals}
\label{subsec:full-t-channel}

\subsection*{Right-handed neutrino}\label{sec:right-hand-neutr-t}

The $t$-channel phase space element for right-handed neutrinos is
given by:
\begin{align}
  \label{eq:lam-t-n-2}
  \Lambda_t^{(N_i)}\left(f_{N_i},f_{q},f_{l_{\alpha}},f_{t}\right)= -
  \frac{ e^{\myE_{l}+\myE_{q}}\,
    \left(-1+f_{N_i}+e^{\myE_{N}}\,f_{N_i}
    \right)}{\left(1+e^{\myE_{l}}\right)\,
    \left(1+e^{\myE_{q}}\right)\,\left(e^{\myE_{l}}+e^{\myE_{l}+\myE_{q}}\right)},
\end{align}
and the $t$-channel contribution of Eq.~(\ref{eq:CI-general-2dim-N})
can equivalently be written as
\begin{equation}
  C_{S,t}[f_{N_i}]= C_{S,t}^{(1)} + C_{S,t}^{(2)} +C_{S,t}^{(3)} +C_{S,t}^{(4)},
\end{equation}
where the constituent integrals are as follows:
\begin{enumerate}
\item First integral (with $\tilde{k}\equiv k/T$):
\begin{align}
  \label{eq:C-t-n-C-1}
  C_{S,t}^{(1)}&=\frac{3\,T}{2^{6}\pi^{3}\, \myE_{N}
    y_{N}}\,\frac{h_t^{2} \,M_i\,\tilde{m}_i}{v^{2}}
  \int_{\frac{1}{2}\left(\myE_{N}-y_{N}+a_h z_i\right)}^{\frac{1}{2}\left(\myE_{N}+y_{N}\right)}
  d\myE_{l} \int_{\frac{1}{2} a_h z_i}
  ^{\frac{1}{2}\left(2\myE_{l}-\myE_{N}+y_{N}\right)} d\myE_{q}\;
  \Lambda_t^{(N)} \; I_{t}^{(1)}, \\ 
\nonumber I_{t}^{(1)} &=
  \int_{\myE_{N}-\myE_{l}+a_h z_i}^{2\myE_{q}+\myE_{N}-\myE_{l}}\,
d \tilde{k} \,
 \frac{\left(\myE_{N}-\myE_{l}\right)^{2}-z^2_i-\tilde{k}^{2}
  }{\left(\myE_{N}-\myE_{l}\right)^{2}
      -\tilde{k}^{2}}   
\\ \nonumber & = \frac{ 2
    \left(\myE_{N}-\myE_{l}\right)\, \left(2\myE_{q}-a_{h}z_i \right)-
    z^2_i\,
    \log\left[\frac{\left(\myE_{N}+\myE_{q}-\myE_{l}\right)\,a_{h}z_i}
      {\myE_{q}\left(2\left(\myE_{N}-\myE_{l}\right)+a_{h}z_i\right)}
    \right]} {2\, \left(\myE_{N}-\myE_{l}\right)}.
   \end{align}

\item Second integral:
\begin{align}
  \label{eq:C-t-n-C-2}
  C_{S,t}^{(2)}&=\frac{3\,T}{2^{6}\pi^{3}\, \myE_{N}
    y_{N}}\,\frac{h_t^{2} \,M_i\,\tilde{m}_i}{v^{2}}
  \int_{\frac{1}{2}\left(\myE_{N}-y_{N} + a_h z_i\right)}^{\frac{1}{2}\left(\myE_{N}+y_{N}\right)}
  d\myE_{l} \int_{\frac{1}{2}(2\myE_{l}-\myE_{N}+y_{N})}^{\infty}
  d\myE_{q} \;\Lambda_t^{(N)} \;I_{t}^{(2)}, \\ \nonumber
  I_{t}^{(2)}&= \int_{\myE_{N}-\myE_{l}+a_h z_i}^{\myE_{l}+y_{N}}\,
d \tilde{k} \,
 \frac{\left(\myE_{N}-\myE_{l}\right)^{2}-z^2_i-\tilde{k}^{2}
  }{\left(\myE_{N}-\myE_{l}\right)^{2}
      -\tilde{k}^{2}}   
\\ \nonumber & =  \frac{2
    \,\left(\myE_{N}-\myE_{l}\right)\,\left(2
      \myE_{l}-\myE_{N}+y_{N}-a_{h}z_i\right) -z^2_i
    \,\left(\log\left[\frac{-a_{h}z_i\,\left(\myE_{N}+y_{N}\right)}
        {\left(\myE_{N}-2\myE_{l}-y_{N}\right)\,
          \left(2\left(\myE_{N}-\myE_{l}\right)+a_{h}z_i\right)} \right]
    \right)} {2\,\left(\myE_{N}-\myE_{l}\right)}.
  \end{align}

\item Third integral:
  \begin{align}
    \label{eq:C-t-n-C-3}
    C_{S,t}^{(3)}&=\frac{3\,T}{2^{6}\pi^{3}\, \myE_{N}
      y_{N}}\,\frac{h_t^{2} \,M_i\,\tilde{m}_i}{v^{2}}
    \int_{\frac{1}{2} \left(\myE_N+y_N\right)}^{\infty} d\myE_{l}
    \int_{\frac{1}{2}\left(2\myE_{l}-\myE_{N}-y_{N} \right)}
    ^{\frac{1}{2}\left(2\myE_{l}-\myE_{N}+y_{N}\right)}
    d\myE_{q}\;\Lambda_t^{(N)} \;I_{t}^{(3)}, \\ \nonumber
    I_{t}^{(3)}&=
    \int_{\myE_{l}-y_{N}}^{2\myE_{q}+\myE_{N}-\myE_{l}}\,
d \tilde{k} \,
 \frac{\left(\myE_{N}-\myE_{l}\right)^{2}-z^2_i-\tilde{k}^{2}
  }{\left(\myE_{N}-\myE_{l}\right)^{2}
      -\tilde{k}^{2}}   
\\ \nonumber & = \frac{2
      \left(\myE_{N}-\myE_{l}\right)\,\left(\myE_{N}+y_{N}+
        2\left(\myE_{q}-\myE_{l}\right) \right) +z^2_i\,\left(
        \log\left[ -\frac{-\myE_{q}\left(\myE_{N}-y_{N}\right)}
          {\left(\myE_{N}+\myE_{q}-\myE_{l}\right)\,
            \left(\myE_{N}-2\myE_{l}+y_{N}\right)}\right]\right)
    }{2\,\left(\myE_{N}-\myE_{l}\right)}.
  \end{align}

\item Fourth integral:
  \begin{align}
    \label{eq:C-t-n-C-4}
    C_{S,t}^{(4)}&= \frac{3\,T}{2^{6}\pi^{3}\, \myE_{N}
      y_{N}}\,\frac{h_t^{2} \,M_i\,\tilde{m}_i}{v^{2}}
    \int_{\frac{1}{2}\left(\myE_{N}+y_{N}\right)}^{\infty} d\myE_{l}
    \int_{\frac{1}{2}\left(\myE_{l}-\myE_{N}+y_{N}\right)}^{\infty}
    d\myE_{q}\;\Lambda_t^{(N)} \;I_{t}^{(4)},\\ \nonumber
    I_{t}^{(4)} &=
    \int_{\myE_{l}-y_{N}}^{\myE_{l}+y_{N}}\,
d \tilde{k} \,
 \frac{\left(\myE_{N}-\myE_{l}\right)^{2}-z^2_i-\tilde{k}^{2}
  }{\left(\myE_{N}-\myE_{l}\right)^{2}
      -\tilde{k}^{2}}   
\\
    \nonumber & = 
    \frac{4\,\left(\myE_{N}-\myE_{l}\right)\,y_{N}+z^2_i\, \log\left[
        \frac{\left(\myE_{N}-y_{N}\right)\,
          \left(\myE_{N}-y_{N}-2\myE_{l}\right)}{\left(\myE_{N}+y_{N}\right)\,
          \left(\myE_{N}+y_{N}-2\myE_{l}\right)} \right]}
    {2\,\left(\myE_{N}-\myE_{l}\right)}.
  \end{align}

\end{enumerate}
In the integrals~(\ref{eq:C-t-n-C-1}) and~(\ref{eq:C-t-n-C-2}) we have
introduced $a_{h}=m_{\Phi}/M_i$ as an infrared cut-off for the
$t$-channel diagram, where $m_{\Phi}$ is the mass of the Higgs boson
which presumably receives contributions from interactions with the
thermal bath, i.e., its value does not correspond to that potentially
measured at the LHC.  The value of $m_\Phi$ can in principle be
deduced from a thermal field theoretic treatment of leptogenesis, and
the analysis of~\cite{Giudice:2003jh} found $m_{\Phi}(T)\simeq
0.4\,T$.  However some open questions still remain and hence in the
present work we prefer to adopt a value of $a_{h}=10^{-5}$, used first
by Luty in~\cite{Luty:1992un}.

\subsection*{Lepton asymmetry}\label{sec:lepton-asymmetry-t}
The $t$-channel phase space element for the lepton asymmetry is given as
\begin{equation}
  \label{eq:lam-t-a-2}
  \Lambda_t^{\left(l-\overline{l}\right)_{\alpha}}
  \left(f_{\left(l-\overline{l}\right)_{\alpha}},f_{t},f_{N_i},f_{q}\right)=
  f_{\left(l-\overline{l}\right)_{\alpha}}\,
  \frac{e^{\myE_{q}}\left(e^{\myE_{l}}\,\left(-1+f_{N_i}\right)
      -e^{\myE_{N}}\,f_{N_i}\right)}
  {\left(1+e^{\myE_{q}}\right)
    \,\left(e^{\myE_{N}}+e^{\myE_{q}+\myE_{l}}\right)}.
\end{equation}

As in Section~\ref{sec:right-hand-neutr-t} we cut-off the integrand in
$C_{S,t}^{(1)}$ and $C_{S.t}^{(2)}$ by adding $a_h$ in the lower
integration
limit of $k$ ( modifying the limits of $\myE_q$ and $\myE_N$ accordingly) .\\
The $t$-channel contribution of the collision
integral~(\ref{eq:CI-general-2dim-asy}) can then equivalently be
written as
\begin{equation}
  C_{S,t}[f_{\left(l-\overline{l}\right)_{\alpha}}]= C_{S,t}^{(1)} + C_{S,t}^{(2)} +C_{S,t}^{(3)} +C_{S,t}^{(4)},
\end{equation}
with:
\begin{enumerate}
\item First integral: 
  \begin{align}
    \label{eq:C-t-l-9}
    C_{S,t}^{(1)}&=\frac{3\,T}{2^{6}\pi^{3}\, \myE_{N}
      y_{N}}\,\frac{h_t^{2} \,M_i\,\tilde{m}_i}{v^{2}}
    \int_{\frac{(2\myE_{l}-a_h z_i)^{2}+z^2_i}{2 (2 \myE_{l}-a_h
        z_i)}}^{\infty} d\myE_{N} \int_{\myE_{N}-\myE_{l}+\frac{1}{2}
      a_h z_i} ^{\frac{1}{2}\left(\myE_{N}+y_{N}\right)}
    d\myE_{q}\; \Lambda_t^{(l-\overline{l})}\;I_{t}^{(1)}, \\
    \nonumber I_{t}^{(1)} &=\int_{\myE_{N}-\myE_{l}+a_h
      z_i}^{2\myE_{q}+\myE_{l}-\myE_{N}}\, d \tilde{k} \,
    \frac{\left(\myE_{N}-\myE_{l}\right)^{2}-z^2_i-\tilde{k}^{2}
    }{\left(\myE_{N}-\myE_{l}\right)^{2} -\tilde{k}^{2}}
    \\
    \nonumber & = -\, \frac{ 2\,\left(\myE_{N}-\myE_{l}\right)\,
      \left(2\,\left(\myE_{N}-\myE_{q}-\myE_{l}\right)+a_{h}z_i\right) +
      z^2_i \,\log
      \left[\frac{\myE_{q}\,a_{h}z_i}{\left(\myE_{q}-\myE_{N}
            +\myE_{l}\right)\,\left(
            2\left(\myE_{N}-\myE_{l}\right)+a_{h}z_i\right)} \right]
    }{2\,\left(\myE_{N}-\myE_{l}\right)}.
  \end{align}

\item Second integral:
  \begin{align}
    \label{eq:C-t-l-11}
    C_{S,t}^{(2)}&=\frac{3\,T}{2^{6}\pi^{3}\, \myE_{N}
      y_{N}}\,\frac{h_t^{2} \,M_i\,\tilde{m}_i}{v^{2}}
    \int_{\frac{(2\myE_{l}-a_h z_i)^{2}+z^2_i}{2 (2 \myE_{l}-a_h
        z_i)}}^{\infty} d\myE_{N}
    \int_{\frac{1}{2}\left(\myE_{N}+y_{N}\right)}^{\infty} d\myE_{q}
    \;\Lambda_t^{(l-\overline{l})}\;I_{t}^{(2)}, \\ \nonumber
    I_{t}^{(2)}&=\int_{\myE_{N}-\myE_{l}+a_h z_i}^{\myE_{l}+y_{N}}\, d
    \tilde{k} \,
    \frac{\left(\myE_{N}-\myE_{l}\right)^{2}-z^2_i-\tilde{k}^{2}
    }{\left(\myE_{N}-\myE_{l}\right)^{2} -\tilde{k}^{2}}
    \\
    \nonumber & = \frac{ 2\,\left(\myE_{N}-\myE_{l}\right)\,
      \left(2\myE_{l}-\myE_{N}+y_{N}-a_{h}z_i\right) - z^2_i \,\log
      \left[\frac{\left(\myE_{N}+y_{N}\right)\,a_{h}z_i}
        {\left(2\myE_{l}-\myE_{N}+y_{N}\right)\,\left(
            2\left(\myE_{N}-\myE_{l}\right)+a_{h}z_i\right)} \right]
    }{2\,\left(\myE_{N}-\myE_{l}\right)}.
  \end{align}

\item Third integral: 
  \begin{align}
    \label{eq:C-t-l-13}
    C_{S,t}^{(3)}&=\frac{3\,T}{2^{6}\pi^{3}\, \myE_{N}
      y_{N}}\,\frac{h_t^{2} \,M_i\,\tilde{m}_i}{v^{2}}
    \int_{z_i}^{\frac{4\myE_{l}^{2}+z^2_i}{4\myE_{l}}} d\myE_{N}
    \int_{\frac{1}{2}(\myE_{N}-y_{N})}^{\frac{1}{2}(\myE_{N}+y_{N})}
    d\myE_{q} \;\Lambda_t^{(l-\overline{l})}\;I_{t}^{(3)},
    \\\nonumber I_{t}^{(3)}&=\int_{\myE_{l}-y_{N}}^{2\myE_{q}+\myE_{l}-\myE_{N}}\, 
d \tilde{k} \,
 \frac{\left(\myE_{N}-\myE_{l}\right)^{2}-z^2_i-\tilde{k}^{2}
  }{\left(\myE_{N}-\myE_{l}\right)^{2}  -\tilde{k}^{2}}
\\
    \nonumber &  = \, \frac{
        2\,\left(\myE_{N}-\myE_{l}\right)\,\left(2\myE_{q}-\myE_{N}+y_{N}\right)
      \, + z^2_i \,\log
        \left[\frac{\left(\myE_{N}-\myE_{q}-\myE_{l}\right)\,
            \left(\myE_{N}-y_{N}
            \right)}{\myE_{q}\,\left(\myE_{N}-2\myE_{l}+y_{N} \right)}
        \right]}{2\,\left(\myE_{N}-\myE_{l}\right)}.
  \end{align}

\item Fourth integral: 
  \begin{align}
    \label{eq:C-t-l-15}
    C_{S,t}^{(4)}&=\frac{3\,T}{2^{6}\pi^{3}\, \myE_{N}
      y_{N}}\,\frac{h_t^{2} \,M_i\,\tilde{m}_i}{v^{2}}
    \int_{z_i}^{\frac{4\myE_{l}^{2}+z^{z_i}}{4\myE_{l}}} d\myE_{N}
    \int_{\frac{1}{2}(\myE_{N}+y_{N})}^{\infty} d\myE_{q}
    \;\Lambda_t^{(l-\overline{l})}\;I_{t}^{(4)}, \\ \nonumber
    I_{t}^{(4)} &=\int_{\myE_{l}-y_{N}}^{\myE_l + y_N}\, 
d \tilde{k} \,
 \frac{\left(\myE_{N}-\myE_{l}\right)^{2}-z^2_i-\tilde{k}^{2}
  }{\left(\myE_{N}-\myE_{l}\right)^{2}  -\tilde{k}^{2}}
\\
    \nonumber&  = \, \frac{
        \,4\,\left(\myE_{N}-\myE_{l}\right)\,y_{N} + z^2_i \,\log
        \left[\frac{\left(\myE_{N}-2\myE_{l}-y_{N}\right)\,\left(\myE_{N}-y_{N}
            \right)}{\left(\myE_{N}-2\myE_{l}+y_{N}
            \right)\,\left(\myE_{N}+y_{N}\right)} \right]
      }{2\,\left(\myE_{N}-\myE_{l}\right)}.
  \end{align}

\end{enumerate}

\section{Reaction rates in the integrated picture}
\label{subsec:reacrates}
The decay interaction term $D_i$ in the Boltzmann
equations~(\ref{eq:BE-Nn-s-1}) and~(\ref{eq:BE-Na-s-1}) is given by
\begin{align}
\label{eq:decayrate}
D_i \equiv z_i\,K_{i}\,\left \langle \frac{M_i}{E_{N}} \right \rangle.
\end{align} 

The scattering rate $S_i$ consists of two terms,
$S_i=2\,S_{s}+4\,S_{t}$, coming respectively from scattering in the
$s$-channel and in the $t$-channel.  One factor of 2 stems from
contributions from processes involving anti-particles, and another
factor of 2 in the $t$-channel term originates from the $u$-channel
diagram.

It is convenient to rewrite the $s$- and $t$-channel scattering rates
$S_{(s,t)_i}$ in terms of the functions $f_{(s,t)_i}$, defined as
\begin{align}
  \label{eq:s-int-fst-z}
  f_{(s,t)_i}(z_i)=\frac{\int_{z^{2}}^{\infty}\,d\varPsi\,\chi_{s,t}\left(\varPsi/z_i^{2}\right)
    \, \sqrt{\varPsi}\,K_{1}\left(\sqrt{\varPsi}\right)}{z_i^{2}\,K_{2}(z_i)},
\end{align}
such that 
\begin{align}
  \label{eq:s-int-Sst}
  S_{(s,t)_i}=\frac{K_{S_i}}{9\,\zeta(3)}\,f_{(s,t)_i},
\end{align}
and the total scattering rate is given by
\begin{align}
  \label{eq:s-int-expl}
  S_i=\frac{2\,K_{S_i}}{9\,\zeta(3)}\,\left(f_{s_i}(z_i)+2\,f_{t_i}(z_i)\right),
\end{align}
where
\begin{align}
  \label{eq:K-s}
  K_{S_i}=\frac{\tilde{m}_i}{m_{\ast}^{S}}, \quad
  m_{\ast}^{S}=\frac{4\pi^{2}}{9}\frac{g_{N}}{h_t^{2}(T)}\,m_{\ast},
\end{align}
with $\tilde{m}_i$ and $m_\ast$ defined in Eqs.~(\ref{eq:mu-nu-eff-eq}).

The
functions $\chi_{s,t}(x)$ are defined as
\begin{align}
  \label{eq:s-int-fs}
  \chi_{s}(x)&=\left(\frac{x-1}{x}\right)^{2},\\
  \label{eq:s-int-ft}
  \chi_{t}(x)&=\frac{x-1}{x}\,\left[\frac{x-2+2a_{h}}{x-1+a_{h}}+\frac{1-2a_{h}}{x-1}\,
    \log\left(\frac{x-1+a_{h}} {a_{h}}\right) \right],
\end{align}
where again $a_{h}=m_{\Phi}/M_i$ is the infrared cut-off for the
$t$-channel diagram.

The total wash-out rate is given by
\begin{align}
    \label{eq:s-int-wo-tot}
    W_i= W_{ID_i}\,\left[1+\frac{1}{D_i} \left(
        2\, \frac{N_{N_i}}{N_{N_i}^{\mathrm{eq}}}\, S_{s_i} +4\,
        S_{t_i}\right)\right]
\end{align}
where $W_{ID_i}$ quantifies the strength of the wash-out due to inverse
decays
\begin{align}
\label{eq:wid}
W_{ID_i}
\equiv\frac{1}{2}\,D_i\,\frac{N^{\mathrm{eq}}_{N_i}}{N^{\mathrm{eq}}_{l}},
\end{align} 
and the equilibrium abundances are
$N_l^{\rm eq}=3/4$, and $N^{\rm eq}_{N_{i}}(z)=\frac{3}{8}
\,z_i^{2}K_{2}\,(z_i)$.

\providecommand{\href}[2]{#2}\begingroup\raggedright\endgroup

\end{document}